\documentclass[a4paper,11pt]{article}
\pdfoutput=1 % if your are submitting a pdflatex (i.e. if you have
\usepackage{jheppub} 

\usepackage[utf8]{inputenc}

\usepackage[T1]{fontenc} % if needed

\usepackage{Extrapackages}
\usepackage{xcolor}
\usepackage{cancel}
\usepackage{tensor}

\usepackage{amsmath}
\newenvironment{rcases}
  {\left.\begin{aligned}}
  {\end{aligned}~\right\rbrace}

\newcommand{\Nf}{{N_{\mathrm{f}}}}

\newcommand{\db}[1]{\dot{\bar{#1}}}

\newcommand{\vv}[1]{\mathbf{#1}}

\newcommand{\eq}[1]{\begin{equation}{#1}\end{equation}}

\definecolor{dmgreen}{rgb}{0.90,0.50,0.10}

\newcommand{\be}{\begin{equation}}
\newcommand{\ee}{\end{equation}}

\title{\boldmath{Hyperinflation generalised:}\\
\Large{from its attractor mechanism to its tension with the `swampland conditions'}}

 \author[1]{Theodor Bjorkmo,}
 \author[2]{M. C. David Marsh}
 \affiliation[1]{Department of Applied Mathematics and Theoretical Physics, \\ University of Cambridge, CB3 0WA Cambridge,  United Kingdom}
  \affiliation[2]{Department of Physics, Stockholm University\\SE - 106 91 Stockholm, Sweden}
  
  \emailAdd{t.bjorkmo@damtp.cam.ac.uk}
  \emailAdd{david.marsh@fysik.su.se}

\arxivnumber{1901.08603}

\abstract{In negatively curved field spaces, inflation can be realised even in steep potentials. Hyperinflation invokes the `centrifugal force' of a field orbiting the hyperbolic plane to sustain inflation. We generalise hyperinflation by showing that it can be realised in models with any number of fields ($N_f\geq2$), and in broad classes of potentials that, in particular, don't need to be rotationally symmetric.  
For example, hyperinflation can follow a period of radial slow-roll inflation that undergoes geometric destabilisation, yet this inflationary phase is not identical to the recently proposed scenario of `side-tracked inflation'.   
We furthermore provide a detailed proof of the attractor mechanism of (the original and generalised)  hyperinflation, and provide a novel set of characteristic, explicit models. We close by discussing the compatibility of hyperinflation with observations and the recently much discussed `swampland conjectures'. Observationally viable models can be realised that satisfy either the `de Sitter conjecture' ($V'/V\gtrsim 1$) or the `distance conjecture' ($\Delta \phi \lesssim 1$), but satisfying both simultaneously brings hyperinflation in some tension with successful reheating after inflation. However, hyperinflation can get much closer to satisfying all of these criteria than standard slow-roll inflation. Furthermore, while the original model is in stark tension with the weak gravity conjecture, generalisations can circumvent this issue.}

\begin{document} 

\maketitle

%%%%%%%%%%%%%%
%%%%%%%%%%%%%

%\emailAdd{t.bjorkmo@damtp.cam.ac.uk}
%\emailAdd{m.c.d.marsh@damtp.cam.ac.uk}

%
\section{Introduction}

%A central goal of modern cosmology is to determine the laws of physics governing the very early universe. 
Inflation is the leading paradigm for explaining the primordial origin of structure in the universe, and rather simple models of inflation are consistent with all current observations. In the standard  slow-roll paradigm involving a single scalar field $\phi$, accelerated expansion is supported when  $\phi$  evolves slowly in a scalar potential potential, $V(\phi)$, that is sufficiently flat:
\be 
\frac{|V_{,\phi}|}{V} < \sqrt{2} \, . \label{eq:SRcond1}
\ee
  Here $V_{,\phi} = \partial_\phi V$ and $\phi$ is taken to have canonical kinetic terms.\footnote{I.e.~${\cal L}/\sqrt{-g} = -\tfrac{1}{2} \partial_{\mu} \phi \partial^{\mu} \phi - V(\phi)$.}
  Throughout this paper we use natural units where the reduced Planck mass is set to one: $M_{\rm Pl} =1/\sqrt{8\pi G} = 2.4\times10^{18}~{\rm GeV} =1$.
    Models of inflation involving multiple fields and non-trivial kinetic terms can also be consistent with observations, and recently, much effort  has been devoted to elucidating the rich possibilities offered by inflationary theories with geometrically non-trivial field spaces (cf.~e.g.~\cite{Kallosh:2013hoa,Kallosh:2013daa,Kallosh:2013yoa,Kallosh:2014rga,Galante:2014ifa,Carrasco:2015uma,Renaux-Petel:2015mga,Linde:2015uga,Carrasco:2015rva,Broy:2015qna,Roest:2015qya, Dias:2015rca,Kallosh:2016sej,Linde:2016uec, Kallosh:2016gqp,Brown:2017osf,Mizuno:2017idt,Kallosh:2017wku,Achucarro:2017ing,Renaux-Petel:2017dia,Linde:2018hmx,Achucarro:2018vey,Christodoulidis:2018qdw,Garcia-Saenz:2018ifx,Garcia-Saenz:2018vqf,Dias:2018pgj,Chen:2018uul,Chen:2018brw,Cremonini:2010ua,Kaiser:2013sna,Baumann:2014nda, Achucarro:2016fby,Karananas:2016kyt}).
%A major theme of these works are that 
The most well-studied class of such models builds on the observation that if the coefficients of the kinetic terms become large, even highly featured  potentials can support standard slow-roll inflation \cite{Kallosh:2013hoa,Kallosh:2013daa,Kallosh:2013yoa,Kallosh:2014rga,Galante:2014ifa,Carrasco:2015uma,Carrasco:2015rva,Broy:2015qna,Roest:2015qya,Kallosh:2016sej,Linde:2016uec, Kallosh:2016gqp,Achucarro:2017ing}. 

However, non-trivial field space geometries can also support conceptually new realisations of inflation, that don't rely on equation \eqref{eq:SRcond1}.
%
%
%
% including the dynamics relevant during inflation. 
% Simple models of single-field slow-roll inflation can be consistent with all observations \cite{}, but inflation may have involved  multiple fields and more general dynamics. Recently, much effort has been devoted to elucidating the rich possibilities offered by inflationary theories with geometrically non-trivial field spaces \cite{}. Such models are well-motivated by ultraviolet completions of inflation into supergravity or string theory, which naturally result in  curved field spaces. 
%
`Hyperinflation' \cite{Brown:2017osf} is a particularly interesting mechanism for realising inflation when the field space is the hyperbolic plane with a constant curvature of $L \ll1$.
Reference \cite{Brown:2017osf} showed that if the field initially has some (field space) angular momentum and the scalar potential is rotationally symmetric, the radial motion  is slowed by an additional centrifugal force, which helps sustain inflation even in steep potentials.\footnote{This mechanism generalises the idea of `spinflation' \cite{Easson:2007dh}.} In this scenario, equation \eqref{eq:SRcond1} is replaced by the much more relaxed condition,
\be
3L < \frac{|V_{,\phi}|}{V} < \frac{1}{L} \, ,
\label{eq:HIcond1}
\ee  
for the canonically normalised radial field $\phi$. 

Reference \cite{Mizuno:2017idt} further developed the perturbation theory of these models, but the framework of hyperinflation is not yet fully developed. In this paper we address the following important questions:
\begin{itemize}
\item The proposal of \cite{Brown:2017osf}, which was further d applies strictly to a restrictive set of  two-field models 
in which the field space is the hyperbolic plane, the scalar potential is rotationally symmetric, and the field is initially displaced far from the origin (so that the initial angular momentum is large).
%Moreover, in \cite{Brown:2017osf} hyperinflation is only realised for  initial conditions with  large  field-space  angular momenta. 
 \emph{Are these assumptions required to realise hyperinflation? If not, how can they be relaxed, and hyperinflation generalised?  }
%\item In \cite{},\emph{How can hyperinflation be realised starting from more general initial conditions, including small or vanishing initial angular momenta?}
\item Models of inflation in steep potentials are potentially interesting because they may not require the same level of tuning as slow-roll models, 
and may be 
 %Moreover, multiple-field models in curved field spaces may be 
 more easily compatible with ultraviolet completions of inflation, e.g.~into string theory. Recently, 
 much activity has been devoted to identifying `swampland' criteria that supposedly delineate the low-energy effective field theories that can be consistently embedded in quantum gravity. In particular, the `de Sitter' and `distance' conjectures respectively state that $|V_{,\phi}| >{\cal O}(V)$ and the total field excursion is bounded from above, $\Delta \phi <{\cal O}(1)$. These conditions, if true, strongly constrain standard slow-roll inflation. \emph{Can hyperinflation be realised in steep potentials with sub-Planckian field excursions? Is the theory compatible with the proposed `swampland conjectures'?} 
\end{itemize}

To address these questions, we identify a set of conditions on the scalar potential and its derivatives that must be satisfied by (generalised) models of hyperinflation. We express these conditions covariantly and show that they can be satisfied in a variety of models, including those without rotational symmetry, with random interactions, and with any number of fields ($N_f\geq2$). Moreover, we demonstrate how (generalised) hyperinflation can be realised without any special conditions on the initial value of the angular momentum (which is not a conserved quantity in the general case). For example,  hyperinflation can follow a period of slow-roll inflation with small or vanishing angular momentum which then becomes `geometrically destabilised' \cite{Renaux-Petel:2015mga} by the negative curvature (cf.~also \cite{Brown:2017osf}). 
This links hyperinflation to so-called `side-tracked inflation' recently discussed in \cite{Renaux-Petel:2015mga, Garcia-Saenz:2018ifx}: we point out that while these setups share some similarities, the side-tracked model of \cite{Garcia-Saenz:2018ifx} does not realise  hyperinflation in either its original or here generalised forms.

We furthermore consider perturbations around the homogeneous solution  and provide the first detailed proof of the attractor nature of hyperinflation, and its generalised versions.
 For models with $N_f>2$ fields, we show that the dynamics of the perturbations decouples the adiabatic and first `entropic' mode from the remaining $N_f-2$ modes. We note that this makes the observational predictions rather independent of the number of fields.

Finally, we discuss the status of hyperinflation relative various conjectures regarding the general properties of effective field theories arising in low-energy limits of consistent theories of quantum gravity, a.k.a.~the `swampland program' \cite{Obied:2018sgi,Ooguri:2018wrx}. We show that hyperinflation can be consistent with either the `de Sitter conjecture' or the `distance conjecture', but struggles to satisfy both conditions simultaneously while also reheating the universe after inflation. Moreover, the simplest models of hyperinflation considered in \cite{Brown:2017osf} are in strong tension with the weak gravity conjecture \cite{ArkaniHamed:2006dz}, but generalisations can be unaffected by this condition.

\section{Hyperbolic geometry and hyperinflation}
In this section, we review the mechanism for hyperinflation proposed in \cite{Brown:2017osf}. To set our notation conventions, we begin with a brief reiew of inflation in curved field spaces.

The action governing inflation in curved field spaces is given by
\eq{
\mathcal S=\frac12\int d^4x\sqrt{-g}\left[ R-G_{ab}g^{\mu\nu}\partial_\mu\phi^a\partial_\nu\phi^b-2V\right],
}
where $G_{ab}$ is the field space metric. In a Friedmann-Robertson-Walker background, the equations of motion for a homogeneous background field become
\eq{
\mathcal D_t\dot\phi^a+3H\dot\phi^a+G^{ab}V_{,b}=0,
}
where the derivative $\mathcal D_t$ is defined by $\mathcal D_tX^a=\dot{X}^a+\Gamma^{a}_{bc}\dot\phi^bX^c$ for any field space vector $X^a$.

The dynamics of the perturbations around a homogeneous solution  in spatially flat gauge is governed by the  second-order action \cite{GrootNibbelink:2001qt,Gordon:2000hv,Sasaki:1995aw}:
\eq{
\mathcal S_2=\frac12\int\frac{d^3k}{(2\pi)^3}dta^3\left[G_{ab}\mathcal D_t\delta\phi^a(\vv k)\mathcal D_t\delta\phi^b(-\vv k)-\left(\frac{k^2}{a^2}G_{ab}+M_{ab}\right)\delta\phi^a(\vv k)\delta\phi^b(-\vv k)\right] \, ,
}
where $M_{ab}$ is the effective mass matrix given by
\eq{
M_{ab}=V_{;ab}-R_{acdb}\dot\phi^c\dot\phi^d+(3-\epsilon)\frac{\dot\phi_a\dot\phi_b}{H^2}+\frac{\dot\phi_aV_{,b}+V_{,a}\dot\phi_b}{H}
}
where $V_{;ab}=V_{,ab}-V_{,c}\Gamma^c_{ab}$. The equations of motion for the perturbations are
\eq{
\mathcal D_t\mathcal D_t\delta\phi^a+3H\mathcal D_t\delta\phi^a+\frac{k^2}{a^2}\delta\phi^a+\tensor{M}{^a_b}\delta\phi^b=0\label{eq:perteq1}.
}

One of the novel aspects of inflation in negatively curved spaces is that the second term can induce a negative mass-squared, leading to geometric destabilisation of inflation attractors \cite{Renaux-Petel:2015mga}\footnote{This destablisation, although very different in origin, is akin to hybrid inflation \cite{Linde:1993cn,GarciaBellido:1996qt}}. In two dimensions, the Riemann tensor always takes the form $\tensor{R}{^a_{bcd}}=R(\tensor{\delta}{^a_c}G_{bd}-\tensor{\delta}{^a_d}G_{bc})/2$, and hyperbolic geometry corresponds to constant $R=-2/L^2$.\footnote{The curvature scale $L$ is related to the eponymous parameter of $\alpha$-attractors by $\alpha = 2L^2/3$.}
 More generally, for an $\Nf$-dimensional hyperbolic geometry, one can show that in an orthonormal basis (normal coordinates), the Riemann tensors takes the form  $\tensor{R}{^a_{bcd}}=-(\tensor{\delta}{^a_c}\delta_{bd}-\tensor{\delta}{^a_d}\delta_{bc})/L^2$. 

Commonly used representations of the metric of the hyperbolic plane include the Poincare disk model,
\be
ds^2 = 4L^2 \frac{dr^2 + r^2 d\theta^2}{(1-r^2)^2}\, , 
\ee
for $r<1$, and the non-compact representation,
\be
ds^2 =d\phi^2 + L^2 \sinh^2(\phi/L) d\theta^2
\, .
\label{eq:Hplane2}
\ee
These are related by the change of radial coordinate $\phi = L \sinh^{-1}(2r/(1-r^2))$.

\subsection{Hyperinflation}
\label{sec:HI}
%Inflation models in hyperbolic field spaces are often studied on the Poincare disc, where the action takes the form
%\eq{
%\mathcal S =\int d^4xa^3\left( 2L^2 \frac{\dot r^2 + r^2\dot\theta^2}{(1-r^2)^2} - V(r,\theta)\right) ,
%}
%and the coordinate $r$ takes values $0\leq r<1$. The with Ricci scalar here is $R=-{2}/{L^2}$, and $L$ is related to $\alpha$ in $\alpha$-attractor models by $L=\sqrt{3\alpha/2}$. This action can be rewritten in terms of a canonically normalised radial coordinate $\phi$ through  $\phi = L \sinh^{-1}(2r/(1-r^2))$, giving 
The action of a homogeneous background field in the hyperbolic plane can be written as
\eq{
\mathcal S=\int d^4xa^3\left(\frac12\dot\phi^2+\frac12L^2\sinh^2(\phi/L)\dot\theta^2-V(\phi, \theta)\right) \, .
}
%and it was in this chart that the original hyperinflation solution was found. 
If the scalar potential only depends on the radial coordinate, $V= V(\phi)$, the equations of motion are given by
\begin{align}
\ddot\phi+3H\dot\phi-L\sinh(\phi/L)\cosh(\phi/L)\dot\theta^2&=-V_{,\phi}\label{eq:eqphi}\\
\ddot\theta+3H\dot\theta+\frac2L\coth(\phi/L)\dot\theta\dot\phi&=0\label{eq:eqtheta} \, .
\end{align}
Following \cite{Brown:2017osf}, we assume that $V(\phi)$ has a minimum at $\phi =0$. For this rotationally symmetric system, equation \eqref{eq:eqtheta} may also be derived as the conservation equation for field-space angular momentum,
$
dJ/dt=0 
$,
where
\eq{
J\equiv a^3L^2\sinh^2(\phi/L)\dot\theta \, .
\label{eq:J}
}
The conservation of $J$ simplifies the radial equation to,
\eq{
\ddot\phi+3H\dot\phi = \frac{J^2\cosh(\phi/L)}{a^6L^3\sinh^3(\phi/L)} -V_{,\phi} \, ,
}
where the first term on the right-hand side may be regarded as a centrifugal force.
%Assuming that the radial acceleration is small, o
One may search for solutions in which $\ddot \phi$ is negligible, and
 the two terms on the right-hand side %are both substantial but cancel with high accuracy. 
 balance to keep $3H \dot \phi$ slowly varying. 
For $\phi \gg L$,  $\sinh(\phi/L)\simeq\cosh(\phi/L)\simeq e^{\phi/L}/2$, and  such solutions must realise the scaling relation
\eq{
a^6 e^{2\phi/L}V_{,\phi}=\text{const} \, .
}
%
%We now want to study a solution with $\ddot\phi=0$ and $J\neq0$ in a region of field space where $\phi\gg L$. These are of course coordinate-dependent statements (as an aside, in these coordinates $J=0$ corresponds to standard slow-roll inflation). We further assume that $3H\dot\phi$ varies slowly, meaning that the two remaining terms must scale the same way with $\phi$. This implies
%where we assume 
%. 
Differentiating this condition with respect to time leads to a simple equation for the radial velocity:
\eq{
\dot\phi
= - \frac{3HL}{1+ \eta_L/2}
\simeq-3HL \, ,\label{eq:hinfphi}
}
where in the last step we have assumed that $|\eta_L| \ll1$ for 
\be
\eta_L = \frac{L V_{,\phi\phi}}{V_{,\phi}} \, .
\ee
The speed in the angular direction may now be found from equation \eqref{eq:eqphi} and is given by
\eq{
L\sinh(\phi/L)\dot\theta\simeq \sqrt{LV_{,\phi}-(3HL)^2} \, .\label{eq:hinftheta}
}
A consistent solution requires $LV_{,\phi} > (3HL)^2$. 

Accelerated expansion is realised if the inflationary epsilon parameter is small:
\eq{
\epsilon=- \frac{\dot H}{H^2} = \frac{\frac12\dot\phi^2+\frac12L^2\sinh^2(\phi/L)\dot\theta^2}{H^2}\simeq3\frac{\tfrac12LV_{,\phi}}{\tfrac12LV_{,\phi}+V} < 1 \, .
\label{eq:HIepsilon}
}
Combining the consistency condition from \eqref{eq:hinftheta} with the inequality \eqref{eq:HIepsilon} leads to equation \eqref{eq:HIcond1}. 
A sustained period of inflation then requires $\epsilon_L \ll 1$ for
\be
\epsilon_L = \frac{L V_{,\phi}}{V} \, .
\ee

%so for inflation to happen we require $LV_{,\phi}\ll V$, as claimed earlier. 
The number of e-folds of expansion is straightforwardly related to the radial field excursion by
\eq{
N=\int Hdt=-\frac1{3L}\int d\phi=\frac{\Delta\phi}{3L}.
}
Obtaining at least $60$ e-folds of inflation requires $\Delta \phi \gtrsim 180 L$. %, which can be sub-Planckian if $L\ll 1$.
The total field excursion  including both the radial and angular motion  is given by
 \be
 d\phi_{\rm tot} = \sqrt{LV_{,\phi}}\, dt = \sqrt{3 \epsilon_L}\, dN \, .
 \ee 
 For future reference we define the  turning-rate parameter of the trajectory by \cite{Brown:2017osf}
\be
\bar h \equiv \frac{L\sinh(\phi/L)\dot\theta}{HL} = 3 \sqrt{\frac{\epsilon_L}{3L^2} -1}
=
\sqrt{\left(\frac{1}{L} \frac{d\phi_{\rm tot}}{dN}\right)^2 -9 }
 \, .
 \label{eq:hbar}
 \ee
We note that the above results come out easily because of the metric choice. In the Poincare disk model, the hyperinflation solution appears more complicated. The large radius condition $\phi\gg L$ clearly corresponds to $r\simeq 1$, but the condition $\ddot\phi\approx0$ translates into the less clear condition
\eq{
\frac{2L}{1-r^2}\left(\ddot r+\frac{2r\dot r^2}{1-r^2}\right)\approx0.
\label{eq:SRpoincare}
}
To obtain a more general understanding of when and how hyperinflation arises, we will now  reformulate the relevant conditions  in a coordinate-independent way. %This is the subject of the next section.

\section{Generalising hyperinflation}
\label{sec:genHI}
In this section we generalise the notion of hyperinflation. As we have reviewed, hyperinflation can be supported in potentials that would be too steep to sustain slow-roll inflation, but the construction of  \cite{Brown:2017osf} required a number of assumptions.  It applied only to two-field models for which the field space was given by the hyperbolic plane and the potential was rotationally symmetric, and it furthermore relied on the initial field being far from the origin with non-vanishing angular momentum, $J\neq 0$. 
There are good reasons to expect that these assumptions can be relaxed: the hyperinflation solution only probes a small region of the field space, and should not be sensitive to global properties of the potential. Moreover, with a general potential there is nothing special about the chosen origin of the hyperbolic plane, and the metric could equally well be expressed locally around a point on the inflationary trajectory.

We begin in section \ref{sec:covar} by reviewing how the Klein-Gordon equation can be expressed covariantly using a local orthonormal basis, and we then show in section \ref{sec:HIgen} how the conditions for hyperinflation can be expressed using only local and covariant properties of the potential. This setup then allows us to generalise hyperinflation to more general potentials in section \ref{sec:Vgen}, and to models with more than two fields in section \ref{sec:Nfgen}.

\subsection{Two-field covariant formulation}
\label{sec:covar}

%In order to generalise hyperinflation we would like to find a coordinate independent description of it and its necessary conditions. In the particular chart used above, the solution is fairly easy to spot, but this does not always have to be the case. Moreover, since a global rotational symmetry is not actually necessary for this solution to arise, we would also like to find a field-space local description of it.

The starting assumptions of the hyperinflation solution of \cite{Brown:2017osf} -- setting $\ddot\phi$ to zero and taking $H\dot\phi$ to be slowly varying -- are not covariant. To generalise hyperinflation we here rephrase the mechanism using vielbeins. A convenient local orthonormal basis of the two-dimensional field space is defined  by the unit vector in the potential gradient direction, 
\be
v^a=V^{,a}/\sqrt{V^{,b}V_{,b}}\equiv V^{,a}/V_{;v}  \, ,
\ee
 and a second unit vector orthogonal to it, $w^a$. The reason for working with this basis, rather than say the commonly used `kinematic basis' that takes $\dot \phi^a$ and $\ddot \phi^a$ to span the field space \cite{Peterson:2010np,Peterson:2011yt}, is that we know the directions of $v^a$ and $w^a$ without solving the equations of motions. This simplifies the analysis. 

In terms of this basis for the tangent space, we expand the field-space velocity as
\eq{
\dot\phi^a=v^a\dot\phi_v+w^a\dot\phi_w \, .
}
Here $\dot\phi_v=v_a\dot\phi^a$ and $\dot\phi_w=w_a\dot\phi^a$ are covariant scalars. As the field evolves, the basis vectors rotate. To take this into account, we note that
%To solve the system, we will need to take into account the rotation of the basis vectors, and we do this by noting that
\eq{
\mathcal D_t v_a=\frac{V_{;ab}\dot\phi^b}{V_{;v}}-v_a\frac{v^bV_{;bc}\dot\phi^c}{V_{;v}}=w_a\frac{w^bV_{;bc}\dot\phi^c}{V_{;v}},
}
where in the last step we used the identity $v^av_b+w^aw_b=\tensor{\delta}{^a_b}$. From the orthonormality condition, $v_aw^a=0$, it also follows that
\eq{
\mathcal D_t w_a=-v_a\frac{w^bV_{;bc}\dot\phi^c}{V_{;v}},
}
In this vielbein basis, the Klein-Gordon equations are then given by
\begin{align}
\ddot\phi_v&=v_a\mathcal D_t(\dot\phi^a)+\mathcal D_t(v_a)\dot\phi^a=-3H\dot\phi_v-V_{;v}+\dot\phi_w\frac{w^aV_{;ab}\dot\phi^b}{V_{;v}} \nonumber \\
\ddot\phi_w&=w_a\mathcal D_t(\dot\phi^a)+\mathcal D_t(w_a)\dot\phi^a=-3H\dot\phi_w-\dot\phi_v\frac{w^aV_{;ab}\dot\phi^b}{V_{;v}} \, . \label{eq:KGeqn1}
\end{align}
These expressions hold in general for two-dimensional field spaces. To identify the local manifestation of the hyperinflation solution, 
%
 %However, to make progress 
 we furthermore need information about the second covariant derivatives of the potential.

\subsection{Generalised hyperinflation}
\label{sec:HIgen}
\label{sec:Vcond}

%To see what the potential should look like locally for hyperinflation, we look again at 
We now reconsider
the hyperinflation solution reviewed in section \ref{sec:HI} and re-express it using our local ON-basis. In the original polar coordinate system, the  basis vectors are $v^a=(1,~0)$ and $w^a=(0,~1/L\sinh(\phi/L))$. Using these, it is straightforward to check that the potential in the original solution satisfies 
\eq{
V_{;v}=V_{,\phi},\hspace{1cm}V_{;vv}=V_{,\phi\phi},\hspace{1cm}V_{;vw}= - \Gamma_{vw}^v V_{,v} =0,\hspace{1cm}V_{;ww}= - \Gamma_{ww}^v V_{,v} = \frac{V_{,\phi}}{L} \, .
}
Here we have introduced the notation $v^aV_{;ab}w^b=V_{;vw}$, etc.~and used the non-vanishing Christoffel symbols $\Gamma^v_{ww} = -  \Gamma^w_{wv} = -\tfrac{1}{L} {\rm cosh}(\phi/L)/{\rm sinh}(\phi/L)$. The hyperinflation parameters $\epsilon_L$ and $\eta_L$ clearly translate into,
\be
\epsilon_L \equiv \frac{LV_{,v}}{V}  \, ,~~{\rm and} ~~\eta_L = \frac{L V_{;vv}}{V_{;v}} \, .
\ee
%Based on this and our earlier discussion, we will therefore
 To generalise hyperinflation, we
 look for solutions of the equation of motion with
 \eq{
|\eta_L| \ll 1 \, ,~~
3L<\epsilon_L<1\, , ~~
V_{;vw} \approx 0 \, ,
~~ {\rm and} ~~
V_{;ww}=\frac{V_{;v}}{L}\label{eq:HIsystem} \, .
}
%We will moreover find that $LV_{;vv}/V_{;v}=\mathcal O(\epsilon)$, and the smallness of $LV_{;vv}/V_{;v}$ is in fact a necessary condition for this solution to arise, even though $V_{;vv}$ does not appear explicitly in the equations for $\ddot\phi_v$ or $\ddot\phi_w$.
%\footnote{We note that these are all 
These conditions are all covariant statements, and can be easily evaluated for any  potential and representation of the metric.

Using the conditions  \eqref{eq:HIsystem}, the basis vector rotations simplify to 
\eq{
\mathcal D_t v_a=w_a\frac{\dot\phi_w}{L},\hspace{1cm}\mathcal D_t w_a= -v_a\frac{\dot\phi_w}{L} \, .
}
%and we consequently obtain t
The equations of motion \eqref{eq:KGeqn1} are now simplified to
\begin{align}
\ddot\phi_v&=-3H\dot\phi_v-V_{;v}+\frac{\dot\phi_w^2}{L} \label{eq:EOM1}\\
\ddot\phi_w&=-3H\dot\phi_w-\frac{\dot\phi_v\dot\phi_w}{L}. \label{eq:EOM2}
\end{align}
Setting $\ddot\phi_v=\ddot\phi_w=0$ at some time $t_p$ gives two possible solutions:
\begin{align}
\begin{rcases}
\db\phi_v&=-V_{v}/3H  \\ 
\db\phi_w&=0
\end{rcases}~\text{slow roll}\, , ~~
~~~
\begin{rcases}
\db\phi_v&=-3HL \\ 
\db\phi_w&= \pm\sqrt{LV_{;v}-9H^2L^2}
\end{rcases}~\text{hyperinflation}
%
%\db\phi_v&= & \db\phi_w&=,
\label{eq:dotbarphi}
\end{align}
%Where we have used the notation $ \dot{\bar{\phi}}_i = \dot \phi_i(t_p)$. 
We here focus on the hyperinflation solution, which again requires $LV_{;v} > 9H^2 L^2$ (or equivalently, $\epsilon_L > 3L$ when $\epsilon_L$ is small). 
It remains to be shown is that these  solutions can be consistently extended along the inflationary trajectory. Noting that the RHS can be expressed entirely as a function of $\phi$ (to a good approximation) during inflation, we look for a solution of the form $\dot\phi=\dot\phi(\phi(t))$, and expand the velocity variables as $\dot\phi_i=\dot{\bar\phi}_i+\delta\dot\phi_i$, where we need to show that $\delta\dot\phi_i(t)\sim {\cal O}(\epsilon\db\phi_i)$ along the inflationary solution.

%We would now like to find a solution satisfying $\dot\phi_{\rm tot}^2\ll V$, and $\ddot\phi_i\sim\epsilon H\dot\phi_i$, ensuring that both $\epsilon$ and $\eta = d\ln \epsilon/dN$ are small, and hence a sustained period of inflation is realised. %The latter condition guarantees $|\mathcal D_t\dot\phi_{\rm tot}^2|=|2\dot\phi_v\ddot\phi_v+2\dot\phi_w\ddot\phi_w|\ll H\dot\phi^2_{\rm tot}$, giving a small $\eta$ with the former.

%To start, we will find the velocities $\db\phi_i(t_p)$ at some time $t_p$ and position $\phi^a(t_p)$, such that $\ddot\phi_i(t_p)$ vanish, giving us $\db\phi_i(t_p)=f_i(\phi(t_p))$. This can be done in any system, but there is no guarrantee that such a solution can be promoted a solution $\db\phi_i(t)=f_i(\phi(t))$ that holds well along the trajectory. To show that this is the case here we find the corrections $\delta\dot\phi_i$ that are needed to make the solution self-consistent along the trajectory, and show that they are indeed of $\mathcal O(\epsilon)$ as claimed earlier.

%the first of which is just standard slow-roll which we will not consider here, and we focus on the second. In this solution, the velocities together satisfy $\db\phi^2=L V_{;v}$. Using these expressions we can compute the inflationary parameters $\epsilon$ and $\eta$. The first, $\epsilon\equiv -\dot H/H^2$, is given by 
When the equations \eqref{eq:dotbarphi} hold, the inflationary slow-roll parameters $\epsilon$ and $\eta$  straightforwardly generalise to
\begin{align}
\epsilon&=\frac{\tfrac12\db\phi^2}{H^2}=3\frac{\epsilon_L}{ \epsilon_L+2} \, ,
\label{eq:epsilonL} \\
\eta &= \frac{d\ln \epsilon}{H dt} = \frac1{H\db\phi^2}\frac{d}{dt}\db\phi^2+2\epsilon=-3\eta_L+2\epsilon \, . \label{eq:etaL}
\end{align}
The assumptions \eqref{eq:HIsystem} imply that both $\epsilon$ and $|\eta|$ are small.
Note that in equation \eqref{eq:etaL}, we take  the time derivative of $\db\phi$ using  the explicit expressions for the $\db\phi_i$ of equation \eqref{eq:dotbarphi},
  rather than the equations of motion (which remain to be solved for $\delta \dot\phi_i$).
%
%so for inflation to happen we must have $\epsilon_L\equiv LV_{;v}/ V\ll1$. The second, $\eta\equiv \dot\epsilon/H\epsilon$, is given by
% From this we see that $|\eta|\ll1$ requires $\eta_L\equiv LV_{;vv}/ V_{;v}\ll 1$, as we assumed earlier. 
%
%If these limits are satisfied we can rewrite the above equations as
%\eq{
%\epsilon_L=\frac{2\epsilon}3,\hspace{1cm}\eta_L=\frac{2\epsilon-\eta}3
%}
%Using these expressions, we can neatly write the $d\db\phi_i/dt$ as
Explicitly,
\eq{
\frac{d\db\phi_v}{dt}=3H^2L\epsilon \hspace{0.5cm}\text{and} \hspace{0.5cm}  \frac{d\db\phi_w}{dt}=\frac{3H(6H^2L^2\epsilon-LV_{;v}\eta_L)}{2\db\phi_w},\label{eq:ddotbarphi}
}
where we emphasise that the right-hand sides can all be written as functions of $\phi^a$ only. The accelerations are both small,
 $d\db\phi_i/dt\sim  {\cal O}( \epsilon H\db\phi_i)$, which justifies the expansion of the velocities. We may now use the full Klein-Gordon equation to  solve for $\delta\dot\phi_i$,
 
% . At first order in $\epsilon$ corrections, we can therefore ignore the , which we expect to be even smaller, on the left hand side of the Klein-Gordon equations. At first order these then become algebraic equations for the 
\eq{
\frac{d\db\phi_v}{dt}=-3H\delta\dot\phi_v+\frac{2\db\phi_w\delta\dot\phi_w}{L},\hspace{1cm} \frac{d\db\phi_w}{dt}=-3H\delta\dot\phi_w-\frac{\db\phi_w\delta\dot\phi_v+\db\phi_v\delta\dot\phi_w}{L} \, .
}
Defining, as in equation \eqref{eq:hbar},  
\be
\bar h= \frac{\db\phi_w}{HL} \, 
\ee
 we find
\eq{
\delta\dot\phi_v=\frac{HL(2\bar h^2\epsilon-(9+\bar h^2)\eta)}{2\bar h^2} \hspace{0.5cm}\text{and} \hspace{0.5cm}  \delta\dot\phi_w=\frac{3HL(\bar h^2(4\epsilon-\eta)-9\eta)}{4\bar h^3} \, .\label{eq:deltadotphi}
}
We note that $\delta\dot\phi_i/\db\phi_i=\mathcal O(\epsilon)$, so that neglecting $d \delta \dot\phi_i/dt \sim \mathcal O(\epsilon^2 \db\phi)$ is consistent. 
%We have now found a self-consistent (perturbative) solution to the equations of motion. By differentiating these expressions with respect to time, we could of course fix the $\mathcal O(\epsilon^2)$ corrections to the velocities, and then repeat the procedure to higher orders, but this is not necessary for our purposes.
This explicitly demonstrates that equation \eqref{eq:dotbarphi} provides a self-consistent solution of the equations of motion to leading order in the slow-roll parameters.

%\tb{[Put this somewhere]} As a quick aside, it is straightforward to see from equation \ref{eq:dotbarphi} that this recovers the original hyperinflation solution.\footnote{Here the potential is rotationally symmetric around the vacuum, so in that particular chart we have $v_a=\left(1,~0\right)$, $w_a=\left(0,~L\sinh(\phi/L)\right)$, and hence $ \dot\phi_v=\dot\phi$, $\dot\phi_w=L\sinh(\phi/L)\dot\theta$.}

As an aside, the equations of motion in hyperinflation resemble those of side-tracked inflation far away from the original axis. In side-tracked inflation \cite{Garcia-Saenz:2018ifx}, the field-space has negative curvature and much like in hyperinflation, slow-roll is destabilised and one ends up in a new attractor solution. However, the equations of motion are not exactly the same, and side-tracked inflation is characterised by $\eta_L=\mathcal O(1)$, and moreover for most of its duration $V_{;vw}$ is non-negligible. One can see some qualitative agreements in the velocities $\dot\phi_v$ and $\dot\phi_w$ between the models, hinting at some connection, but these two inflation models are clearly not the same.

\subsection{Hyperinflation in random potentials}
\label{sec:Vgen}
The conditions on the derivatives of the potential in equation \eqref{eq:HIsystem} are unusual, and may seem highly restrictive. In this section, we show that they are satisfied by broad classes of potentials, and do not rely on rotational symmetry. 

For concreteness, we consider a random potential over $\mathbb R^2$ with derivatives $\partial_x V \sim \partial_y V$.  In a polar coordinate system, this corresponds to
\eq{
|V_{,\phi}|\sim| V_{,\theta}|/\phi,\hspace{1cm}|V_{,\phi\phi}|\sim |V_{,\phi\theta}|/\phi\sim |V_{,\theta\theta}|/\phi^2 \, .\label{eq:R2}
}
When considered as a potential over the hyperbolic plane with a metric
of the form of equation \eqref{eq:Hplane2}, any angular  gradients flatten exponentially when $\phi\gg L$. Despite the assumed randomness, $V_{;v} \approx V_{\phi}$ at large radius. The condition on the steepness of the potential from equation \eqref{eq:HIsystem} (and the earlier equation \eqref{eq:HIcond1}) forces the potential  to be asymptotically  intermediate between the two exponential functions,
\be
{\rm exp}(3\phi L/M_{\rm Pl}^2)~~~\text{and}~~~{\rm exp}(\phi/L) \,,
\ee
where we for clarity have re-inserted factors of the reduced Planck mass. We note that over  a field displacement of $\Delta \phi = 180L$, the potential may change by as little as ${\rm exp}( (23L)^2)$ or as much as ${\rm exp}(180)$. We will return to this point in section \ref{sec:swampland}. 
%
%
%satisfied by a broad class of potentials that are not rotationally symmetric. 
%
%
%Hyperinflation was originally proposed in a rotationally symmetric potential. These potentials seem to imply translational symmetry over distances $L\sinh(\phi/L)$, which for $\phi\gg L$ may well be trans-Planckian. However, the distance in this direction traversed during inflation is of order $\Delta N\bar h L$, which for finite $h$ and small $L$ can easily be sub-Planckian. We therefore do not actually require the global symmetry that the initial model proposed. Having a (small) section of the potential that is well-described by this is sufficient for hyperinflation.
%
%More generally, the hyperinflation attractor solution arises where $LV_{;v}>9H^2L^2$ and the Hessian is well-approximated by equation \ref{eq:HIsystem}. We would now like to discuss when these conditions are well-satisfied in the context of a metric 
 %and potential $V=V(\phi,\theta)$.

At large radius, the Hessian of the random potentials also takes a simple form, with coefficients that satisfy equation \eqref{eq:HIsystem}:
%It is easy to see that the original solution (the one obtained from the Euler-Lagrange equations) remains more or less the same even in the with an angular dependence of the potential, as long as it satisfies
%\eq{
%|V_\theta\ll| L\sinh(\phi/L)|V_{,\phi}|, \hspace{0.33cm}|V_{\theta\phi}|\ll L\sinh(\phi/L)|V_{,\phi\phi}|, \hspace{0.33cm}|V_{\theta\theta}|\ll L^2\sinh^2(\phi/L)|V_{,\phi\phi}|, \label{eq:derivativeconditions}
%}
%etc. If these conditions are satisfied, we also find
\eq{
%V_{;v}\simeq V_{,\phi},\hspace{0.5cm}
V_{;vv}\simeq V_{,\phi\phi},\hspace{0.5cm}V_{;vw}\simeq0, \hspace{0.5cm}\text{and}\hspace{0.5cm}V_{;ww}\simeq\frac{V_{;v}}L \, .
}
Thus, hyperinflation draws heavily on the negative curvature of the hyperbolic plane, but is largely insensitive to the precise form of the potential.

\subsection{Multifield generalisation}
\label{sec:Nfgen}
\label{sec:multifield}

Our formalism also allows us to straightforwardly consider the case of hyperinflation in negatively curved field spaces with $\Nf>2$ dimensions. With $\Nf$ fields, we solve the equations of motion in the vielbein orthonormal basis $\tensor{e}{^I_a}=\{v_a,~\tensor{w}{^1_a},~\tensor{w}{^2_a},...\}$. We will be more specific on the definition of  $\tensor{w}{^1_a}$ below. Following our discussion in section \ref{sec:Vcond}, we consider scalar potentials for which the Hessian  in this local basis takes the diagonal form $V_{;ab}=\text{diag}(V_{;vv},~V_{;v}/L,~V_{;v}/L,\dots)$.\footnote{This, for example, happens for if one takes a radial potential in $\mathbb H^3$, or at large $\phi$ for metrics of the form $ds_{\Nf}^2 = d\phi^2+L^2 \sinh^2(\phi/L) d\Omega_{\Nf-1}^2$.}

Here we study the evolution of $\dot\phi^a$ by looking at its component in the gradient-direction, $\dot\phi_v$, and its complement, $\dot\phi_\perp^a$. To proceed we first note that
\eq{
\mathcal D_tv^a=(g^{ab}-v^av^b)\frac{V_{;bc}\dot\phi^c}{V_{;v}}\equiv \tensor\perp{^a^b}\frac{V_{;bc}\dot\phi^c}{V_{;v}},
}
where $\tensor\perp{^a_b}$ is a projection tensor that removes the component in the gradient direction. The Klein-Gordon equation for the homogenous background field then implies that
\begin{align}
\ddot\phi_v&=\mathcal D_t(v_a\dot\phi^a)=-3H\dot\phi_v-V_{;v}+\frac{\dot\phi^a_\perp V_{;ab}\dot\phi^b}{V_{;v}} \, ,\\
\ddot\phi_\perp^a&=\mathcal D_t(\tensor{\perp}{^a_b}\dot\phi^b)=-3H\dot\phi_\perp^a-\tensor\perp{^a^b}\frac{V_{;bc}\dot\phi^c}{V_{;v}}\dot\phi_v-v^a\frac{\dot\phi^b_\perp V_{;bc}\dot\phi^c}{V_{;v}} \, .
\end{align}
The rate of change of the magnitude of $\dot\phi_\perp^a$ is then given by
\eq{
\ddot\phi_\perp=\frac1{\dot\phi_\perp}\dot\phi_{\perp a}\mathcal D_t\dot\phi_\perp^a=-3H\dot\phi_\perp-\frac{\dot\phi^a_\perp V_{;ab}\dot\phi^b}{\dot\phi_\perp V_{;v}}.
}

Now, in the case of hyperinflation, we have $V_{;ab}=v_av_b V_{;vv}+\perp_{ab}V_{;v}/L$, and hence
\begin{align}
\ddot\phi_v&=-3H\dot\phi_v-V_{;v}+\frac{\dot\phi^2_\perp}L 
\label{eq:multifieldv}
\\
\ddot\phi_\perp^a&=-3H\dot\phi_\perp^a-\frac{\dot\phi_v\dot\phi^a_\perp}{L}-v^a\frac{\dot\phi^2_\perp}{L}.
\end{align}
The acceleration $\ddot\phi_\perp^a$ is therefore in the plane spanned by $v^a$ and $\dot\phi^a_\perp$, and the motion is planar. It therefore suffices to look at the evolution of the norm $\dot\phi_\perp$, which is governed by the equation: 
\eq{
\ddot\phi_\perp=-3H\dot\phi_\perp-\frac{\dot\phi_v\dot\phi_\perp}{L}.
\label{eq:multifieldw}
} 
We note that equations \eqref{eq:multifieldv} and \eqref{eq:multifieldw} are identical to the two-field equations \eqref{eq:EOM1} and \eqref{eq:EOM2} (with  $\dot\phi_\perp \to\dot\phi_w$). Importantly, this means that the most important dynamics of the $\Nf$-dimensional system reduces to the standard two-field hyperinflation equations, with  $\Nf-2$ perpendicular directions decoupled.\footnote{In view of section \ref{sec:Vgen}, we expect that multifield hyperinflation can be realised in sufficiently steep random potentials. This could be tested explicitly by adapting the methods of e.g.~\cite{Marsh:2013qca, Dias:2016slx,Dias:2017gva,Bjorkmo:2017nzd,Bjorkmo:2018txh}. } We will now turn to the dynamics of long-wavelength perturbations around this system, where we will in particular show that this decoupling of modes persists.

\section{Linear stability}
In this short section, we provide the first detailed proof of the attractor nature of (generalised) hyperinflation by showing that small perturbations away from the solution decay. We begin by considering the two-field case, which then straightforwardly generalises to the general case with $\Nf\geq2$ fields.  

\subsection{The two-field case}
To prove the linear stability of this solution, we look at the equations for the $k=0$ mode of linearised perturbations around the background solution:
\eq{
\mathcal D_t\mathcal D_t\delta\phi^a+3H\mathcal D_t\delta\phi^a+[(k^2/a^2H^2)\tensor{\delta}{^a_b}+\tensor{M}{^a_b}]\delta\phi^b=0\label{eq:perteq2}.
}
During hyperinflation, the parameters  $\epsilon$ and $\eta$ are small, and we here work to first order in these parameters. We will prove that $k=0$ perturbations around the background solution decay unless they are in a phase-space direction that still satisfies the hyperinflation equation of motion.

To study the perturbations, it's convenient to use a kinematic basis with  $n^a=\dot\phi/|\dot\phi|$ and $s^a=\mathcal D_t n^a/|\mathcal D_t n^b|$ \cite{Achucarro:2010jv,Achucarro:2010da, GrootNibbelink:2001qt,Peterson:2010np,Peterson:2011yt} (cf.~Appendix \ref{appendix:stability} for more details). The turn-rate of the field is given by
\eq{
a^a\equiv\mathcal D_N n^a=\frac{\dot\phi_wV_{v}}{H\dot\phi^2}s^a=\left(h-\frac{3\eta}{h}\right)s^a\equiv \omega s^a \, ,\label{eq:rotation}
}
where $h=\dot\phi_w/HL$. In this basis, the equations of motion for the $k=0$ mode are given by
%To do this we will make further use of the mathematical framework of vielbeins, which in two dimensions in particular gives us a very convenient framework for looking at the perturbations. The techniques used were first developed {\color{red} by Achucarro et al. \cite{}}, and details are given in . Taking the usual definitions, and working in the corresponding vielbein basis where $\delta\phi^I=(n_a\delta\phi^a,~s_a\delta\phi^a$), we show that the 
%\begin{align*}
%\frac{d^2\delta\phi^n}{dN^2}+(3-\epsilon)\frac{d\delta\phi^n}{dN}-\left[2h-\frac{3\eta}h\right]\frac{d\delta\phi^s}{dN}-\frac{3\eta}2\delta\phi^n-h(6-2\epsilon+2\eta)\delta\phi^s&=0\\
%\frac{d^2\delta\phi^s}{dN^2}+(3-\epsilon)\frac{d\delta\phi^s}{dN}+\left[2h-\frac{3\eta}h\right]\frac{d\delta\phi^n}{dN}-h\eta\delta\phi^n-\left[\frac13h^2(6-2\epsilon+\eta)-\frac{9\eta}{2}\right]\delta\phi^s&=0\,.
%\end{align*}
\begin{align}
\frac{d^2\delta\phi^n}{dN^2}+(3-\epsilon)\frac{d\delta\phi^n}{dN}-2\omega\frac{d\delta\phi^s}{dN}-\frac{3\eta}2\delta\phi^n-\left[\omega(6-2\epsilon+\eta)+\frac{(9+\omega^2)\eta}\omega\right]\delta\phi^s&=0\\
\frac{d^2\delta\phi^s}{dN^2}+(3-\epsilon)\frac{d\delta\phi^s}{dN}+2\omega\frac{d\delta\phi^n}{dN}-\omega\eta\delta\phi^n-\left[\frac13\omega^2(6-2\epsilon+\eta)+\frac{3(9+\omega^2)\eta}{\omega^2}\right]\delta\phi^s&=0\,.
\end{align}
 Defining $\delta\pi^I\equiv d\delta\phi^I/dN$, we can rewrite this system as
\eq{
\frac{d}{dN}\begin{pmatrix}\delta\phi_n \\ \delta\phi_s \\ \delta\pi_n \\ \delta\pi_s\end{pmatrix}=
\begin{pmatrix}0 & 0 & 1 &0\\
0 & 0 & 0 &1\\
3\eta/2, & \omega(6-2\epsilon+\eta)+\frac{(9+\omega^2)\eta}\omega ,&-(3-\epsilon) & 2\omega\\
\omega\eta, & \frac13\omega^2(6-2\epsilon+\eta)+\frac{3(9+\omega^2)\eta}{\omega^2},& -2\omega & -(3-\epsilon)
\end{pmatrix}
 \begin{pmatrix}\delta\phi_n \\ \delta\phi_s \\ \delta\pi_n \\ \delta\pi_s\end{pmatrix}.
 \label{eq:pertmatrixeqn}
}
%In this formulation, taking 
We assume  that 
$\epsilon$ and $\eta$ vary slowly during inflation so that they can  consistently be treated as constant to linear order.  Local  stability of the system can now be determined by computing  the local Lyapunov exponents, i.e.~the eigenvalues $\lambda$ of the evolution matrix of equation \eqref{eq:pertmatrixeqn}. To leading order, they are given by
\begin{align}
\lambda_1&=\eta/2+\mathcal O(\epsilon^2)\, , \\
\lambda_2 &=-3+\mathcal O(\epsilon) \, , \\
\lambda_{3,4} &= \frac12\left(-3\pm\sqrt{9-8\omega^2}\right)+\mathcal O(\epsilon) \, .
\end{align}
The latter three all have negative real parts and thus correspond to decaying modes. The first one is a near-constant mode corresponding to the direction $(1,0,\eta/2,0)^T$. In the $(n^a, s^a)$-basis, this mode is given by,
\eq{
\delta\phi^I\propto(\sqrt{\epsilon},0) \, ,\label{eq:constantmode2}
}
%These perturbations correspond to shifts which still satisfy the constraint.
and is the regular superhorizon-scale adiabatic mode corresponding to shifts along the hyperinflation solution.\footnote{It follows from equation \eqref{eq:constantmode2} that the curvature perturbation in constant density gauge freezes out on superhorizon scales.} Since all other modes decay, we confirm that the general two-field  hyperinflation  solutions are stable.

%This result also shows us that the curvature perturbation does indeed (at least eventually) freeze out on superhorizon scales.
 We note in closing that, due to the high turning rate of the background-solution, the adiabatic mode is  a little bit different from the standard slow-roll case. For a perturbation $\delta\phi_\text{ad}=A(1,~0)$ with $dA/dN=\eta A/2$, the momentum perturbation is hyperinflation given by $\mathcal D_N\delta\phi_\text{ad}=A (\eta /2,~\bar h)$, and is therefore of comparable size to the field perturbation, and in some limits much larger. By contrast, in standard slow-roll inflation, $\mathcal D_N\delta\phi_\text{ad}=\eta\delta\phi_\text{ad}/2$, which is suppressed relative the field perturbation.

\subsection{Perturbations in the general multifield case}
We now show that the decoupling we identified in the background equations of generalised hyperinflation with $\Nf$ fields (cf.~section \ref{sec:multifield})  applies to the perturbations as well. This enables us to reduce the general  problem of the perturbations to the two-field case discussed above. 

%To study the perturbations we now need to have a closer look at our basis vectors. Since the motion is planar and $\dot\phi_\perp\neq0$ 

We can without loss of generality set $\tensor{w}{_1^a}\propto \tensor{\perp}{^a_b}\dot\phi^b$. It then follows that
\eq{
\mathcal D_tv^a=\frac{\dot\phi_{w_1}}L\tensor{w}{_1^a},\hspace{1cm}\mathcal D_t\tensor{w}{_1^a}=-\frac{\dot\phi_{w_1}}Lv^a. \label{eq:mfvw1}
}
These two vectors rotate in their own plane, meaning that the other basis vectors must rotate in the perpendicular subspace that they span. If we have only one more vector it follows automatically that its time derivative must vanish, but if we have more, it will be basis-dependent. 

To proceed we now need to define a kinematic basis. We write it as $\tensor{e}{^I_a}=\{n_a,~\tensor{s}{^1_a},...\}$, where we take $n^a=(\dot\phi_vv^a+\dot\phi_{w_1}\tensor{w}{_1^a})/\dot\phi$ and $\tensor{s}{_1^a}=(-\dot\phi_{w_1}v^a+\dot\phi_{v}\tensor{w}{_1^a})/\dot\phi$. The evolution equations for the perturbations in the $n^a$ and $\tensor{s}{_1^a}$ directions 
are again given by 
equation \eqref{eq:EOM1} and \eqref{eq:EOM2}, and the solution in this plane remains an attractor solution. We do not have explicit expressions for the remaining basis vectors, 
but since the mass matrix for these perturbations is proportional to the identity matrix, one can show (see Appendix \ref{appendix:stability}) that it is always possible to pick a basis for the perturbations where the equations of motion take the form
\eq{
\frac{d^2\delta\phi^{s_I}}{dN^2}+(3-\epsilon)\frac{d\delta\phi^{s_I}}{dN}+\left[\frac{k^2}{a^2H^2}-\frac{3(9+\omega^2)\eta}{2\omega^2}\right]\delta\phi^{s_I}=0 \, .
\label{eq:iso}
}
These are all uncoupled near-massless modes that don't source the adiabatic perturbations. In section \ref{sec:obsgen} we will argue that even if these entropic modes are numerous, the level of isocurvature is naturally suppressed.

%In this section we discuss what combinations of potentials and metrics give rise to hyperinflation as discussed above, and also show that hyperinflation can be extended to include more than two-fields. 

\section{Examples}
In this section we provide several explicit examples of models of hyperinflation that in various ways generalise the simple models discussed in \cite{Brown:2017osf, Mizuno:2017idt}. In particular, our examples demonstrate that hyperinflation can proceed in highly asymmetric scalar potentials, and may follow a period of  slow-roll inflation that becomes geometrically destabilised by the negative curvature. We furthermore present the first explicit models with a sub-Planckian field excursion, and models with more than two fields. Throughout this section, we concentrate on the problem of realising inflation; we will turn to the observational predictions  in section \ref{sec:swampland}.

\subsection{Example 1: hyperinflation in the Poincare patch}
In section \ref{sec:HI} we noted that the conditions under which hyperinflation is realised are non-trivial in the Poincare patch, cf.~equation \eqref{eq:SRpoincare}. The formalism we have developed in section \ref{sec:genHI} makes this issue much more clear. Assuming a radial potential, the vielbein are given by $v^a=((1-r^2)/2L,0)$ and $w^a=(0,(1-r^2)/2Lr)$, and the velocities in the hyperinflation solution are given by,
\eq{
\dot\phi_v=\frac{2L}{1-r^2}\dot r,\hspace{1cm}\dot\phi_w=\frac{2Lr}{1-r^2}\dot\theta.
\label{eq:HIpoincare2}
}
Differentiating the first equation of \eqref{eq:HIpoincare2} and setting the acceleration to zero gives the condition \eqref{eq:SRpoincare}. 

The class of potentials that support hyperinflation in the Poincare patch need to be sufficiently steep close the boundary of the Poincare disk. In the original coordinates, the velocities are given by
\eq{
\dot r=-\frac32 H(1-r^2),\hspace{1cm}\dot \theta=\frac{1-r^2}{2Lr}\sqrt{\frac{1-r^2}2V_{,r}-9H^2L^2}.
}
%Suppose we want to have hyperinflation near the boundary. Taking
 Near the boundary it's convenient to express $r\equiv1-\delta$, and the condition $3L < \epsilon_L <1$ translates into:
\eq{
\frac{3L^2}{\delta}< \frac{V_{,r}}{V} < \frac{1}{\delta} \, .
}
Thus, hyperinflation is supported by a class of potentials that grow sufficiently rapidly towards the edge of the Poincare disk.\footnote{For less rapidly growing potentials, $\alpha$-attractor slow-roll inflation becomes possible. }

%The right-hand side is positive and diverges as $\delta\to0$, and hence this potential must also diverge as $r\to1$. This was expected, considering the coordinate transform $\phi = L \sinh^{-1}(2r/(1-r^2))$, where we see that $\phi$ diverges as $r\to1$. While the divergent quantities makes this chart inconvenient to use for hyperinflation, the radial coordiante does have a neat evolution near the boundary: to leading order it satisfies $\dot\delta=3H\delta$. 

\subsection{Examples 2 \& 3: the non-symmetric hyperinflation attractor and `geometric destabilisation' \label{sec:nonsymmetric}}
The original model of hyperinflation considered rotationally symmetric potentials so that the field space angular momentum (cf.~equation \eqref{eq:J}) is conserved. We now  give two examples of hyperinflation realised in non-symmetric potentials, to show explicitly that this assumptions can be relaxed. These examples also demonstrate how hyperinflation can follow `geometric destabilisation' \cite{Renaux-Petel:2015mga} of slow-roll inflation. \\

\noindent {\bf Example 2:} As a first example of a non-symmetric hyperinflation, we consider the model
\eq{
G_{ab}= \left(
\begin{array}{cc}
 \frac{2 \cosh ^2\left(\frac{\phi}{L}\right) \cosh ^2\left(\frac{\chi}{L}\right)}{\cosh \left(\frac{2 \phi}{L}\right)+\cosh \left(\frac{2 \chi}{L}\right)} & -\frac{\sinh \left(\frac{2 \phi}{L}\right) \sinh \left(\frac{2 \chi}{L}\right)}{2 \left(\cosh \left(\frac{2 \phi}{L}\right)+\cosh \left(\frac{2 \chi}{L}\right)\right)} \\
 -\frac{\sinh \left(\frac{2 \phi}{L}\right) \sinh \left(\frac{2 \chi}{L}\right)}{2 \left(\cosh \left(\frac{2 \phi}{L}\right)+\cosh \left(\frac{2 \chi}{L}\right)\right)} & \frac{2 \cosh ^2\left(\frac{\phi}{L}\right) \cosh ^2\left(\frac{\chi}{L}\right)}{\cosh \left(\frac{2 \phi}{L}\right)+\cosh \left(\frac{2 \chi}{L}\right)} \\
\end{array}
\right),\hspace{0.5cm}V=\frac12 m^2\phi^2.
}
This metric is constructed such that on the $\phi$ and $\chi$ axes, the metric becomes the identity matrix. The curvature is again negative and constant: $R= -2/L^2$.  The quadratic Lagrangian for homogeneous perturbations around an assumed background solution on the $\phi$-axis takes the form
\eq{
\mathcal L\Big|_{\bar \chi =0}=\frac12a^3\left[\delta\dot\phi^2+\delta\dot\chi^2-\left(V_{,\phi\phi}+\frac{2V_{,\phi}\dot\phi}{H}+(3-\epsilon)\dot\phi^2\right)\delta\phi^2-\left(\frac{LV_{,\phi}-\dot\phi^2}{L^2}\right)\delta\chi^2\right] \, .
}
The negative term in the effective $\chi$-mass, $-\dot\phi^2/L^2$, is identical to the term appearing in \cite{Renaux-Petel:2015mga}, where it was identified as the trigger of geometric destabilisation. The slow-roll solution, $\dot\phi=-V_{,\phi}/3H$ is stable as long as $V_{,\phi}<9H^2L=3V L$. Once the gradient exceeds this value, slow-roll undergoes geometric destabilisation, and the solution transitions into the hyperinflation attractor.

\begin{figure}
    \centering
    \begin{subfigure}{0.48\textwidth}
         \includegraphics[width=1\textwidth]{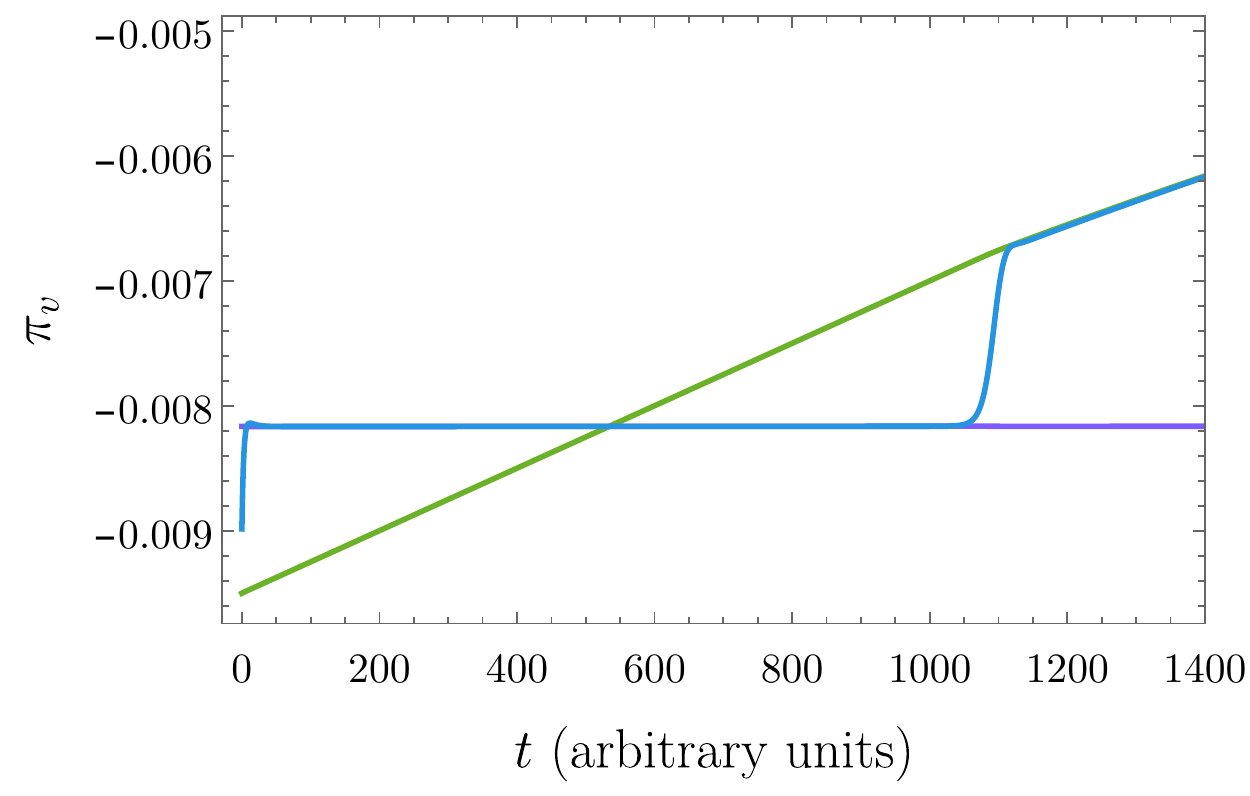}
   % \caption{Fraction of negative eigenvalues}
%    \label{fig:diso}
    \end{subfigure}
%    \caption{These graphs show the effect of varying the }\label{fig:dSHiso}
%\end{figure}
%
%\begin{figure}
%    \centering
~
    \begin{subfigure}{0.48\textwidth}
         \includegraphics[width=1\textwidth]{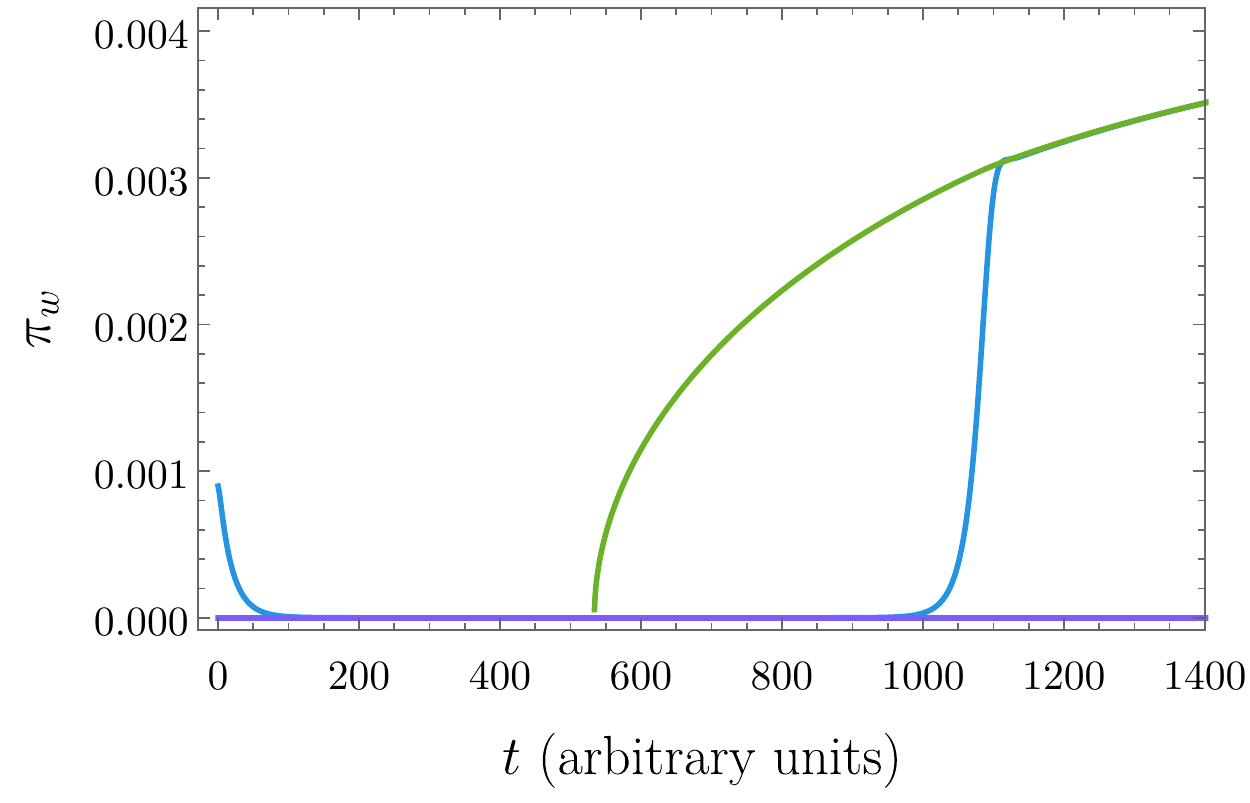}
   % \caption{e-folds before end when a second direction becoms tachyonic}
%    \label{fig:dns}
    \end{subfigure}
     %add desired spacing between images, e. g. ~, \quad, \qquad, \hfill etc. 
      %(or a blank line to force the subfigure onto a new line) \\
	    \caption{The analytical predictions from hyperinflation (green) and slow-roll inflation (purple) together with a numerical realisation of Example 2 (blue).
	    The momenta are quickly attracted to the slow-roll solution, but then destabilise and approach the hyperinflation attractor. \label{fig:nsymmHI}}
\end{figure}

In our example, with $V=\frac12m^2\phi^2$, the instability kicks in at $\phi=2/(3L)$. Further out than this,  slow-roll remains the attractor. In the case shown in Figure \ref{fig:nsymmHI}, which has $L=0.05$, we start at $\phi=31$ with initial momenta  ($\pi_v\equiv\dot\phi_v$ and $\pi_w\equiv\dot\phi_w$) perturbed away from the slow-roll solution. The figure shows how the momenta quickly converge back to the slow-roll solution, but eventually as the fields pass the instability point, they transition into the hyperinflation attractor.

A Planck-compatible model with this metric can easily be constructed with an exponential potential, $V_0=V_0e^{\phi/\lambda}$, and fields starting on the $\phi$ axis. Models with exponetial potentials can be shown to have $\eta\approx 0$, giving $n_s-1=-2\epsilon$ \cite{Mizuno:2017idt}. A Planck compatible spectral index is therefore obtained by setting $\lambda=3L/(1-n_s)$. 
\\

\noindent {\bf Example 3:} We now consider the model defined by the metric and the non-rotationally symmetric potential, 
\eq{
G_{ab}= \text{diag}\left(1,L^2\sinh^2(\phi/L) \right),\hspace{0.5cm}V=\frac12 m^2\phi^2\cos^2\theta \label{eq:nsymm2} \, .
}
At $\phi\gg L$, $V_{;v}=m^2\phi\cos^2\theta=2V/\phi$, so slow-roll undergoes geometric destabilisation at $\phi=2/(3L)$ for all angles (except $\theta=\pi,3\pi/2)$). Hyperinflation is only possible inside of this radius, and until $\phi=\mathcal O(L)$. This metric is easy to work with everywhere in the hyperbolic plane, and here we explore more general initial conditions.

\begin{figure}
    \centering
    \begin{subfigure}{0.4\textwidth}
         \includegraphics[width=1\textwidth]{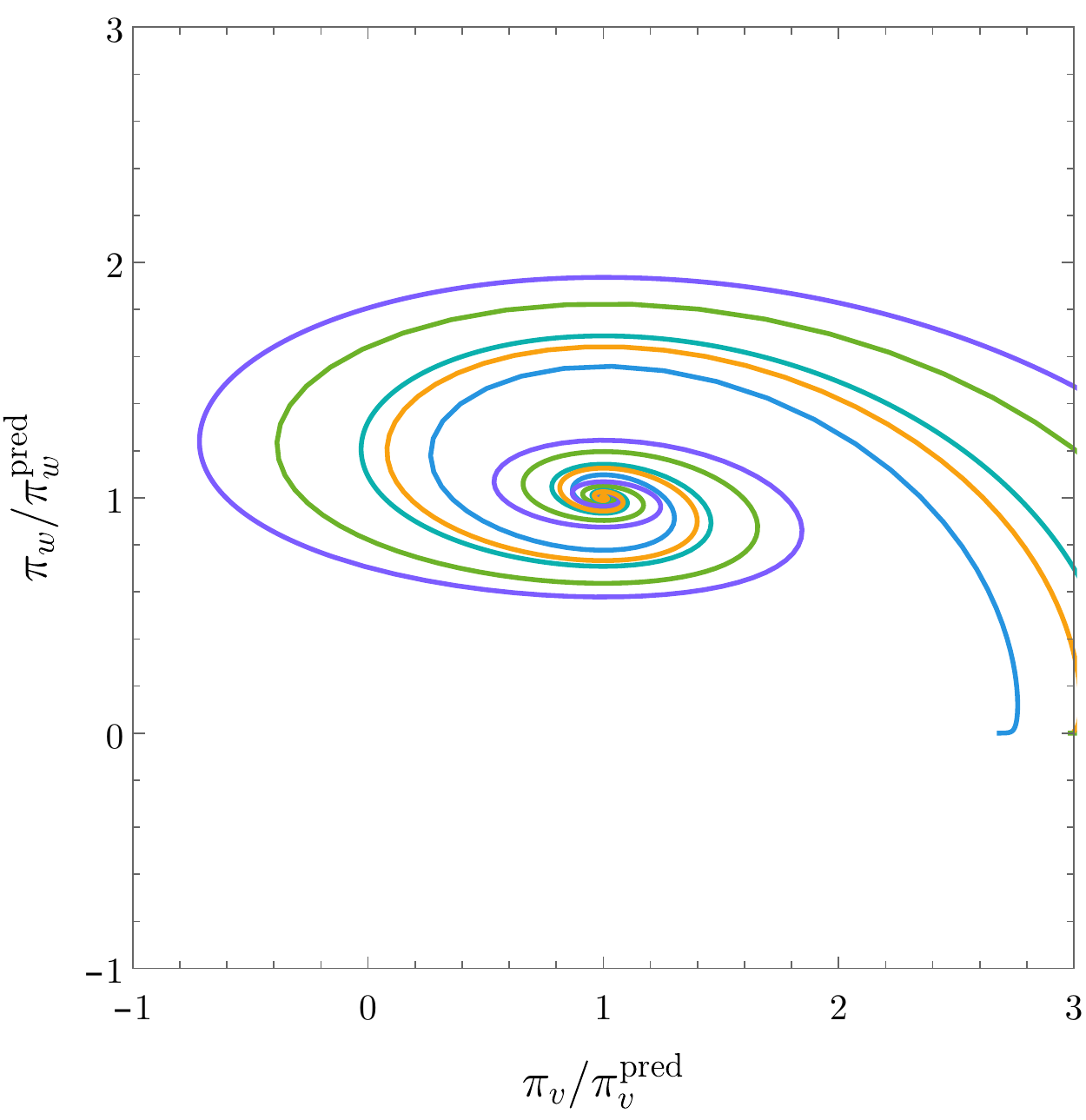}
   % \caption{Fraction of negative eigenvalues}
%    \label{fig:diso}
    \end{subfigure}
%    \caption{These graphs show the effect of varying the }\label{fig:dSHiso}
%\end{figure}
%
%\begin{figure}
%    \centering
%~
%%
%    \begin{subfigure}{0.48\textwidth}
%         \includegraphics[width=1\textwidth]{example1convergence.pdf}
%   % \caption{e-folds before end when a second direction becoms tachyonic}
%%    \label{fig:dns}
%    \end{subfigure}
%     %add desired spacing between images, e. g. ~, \quad, \qquad, \hfill etc. 
%      %(or a blank line to force the subfigure onto a new line) \\
      \\
          \begin{subfigure}{0.48\textwidth}
         \includegraphics[width=1\textwidth]{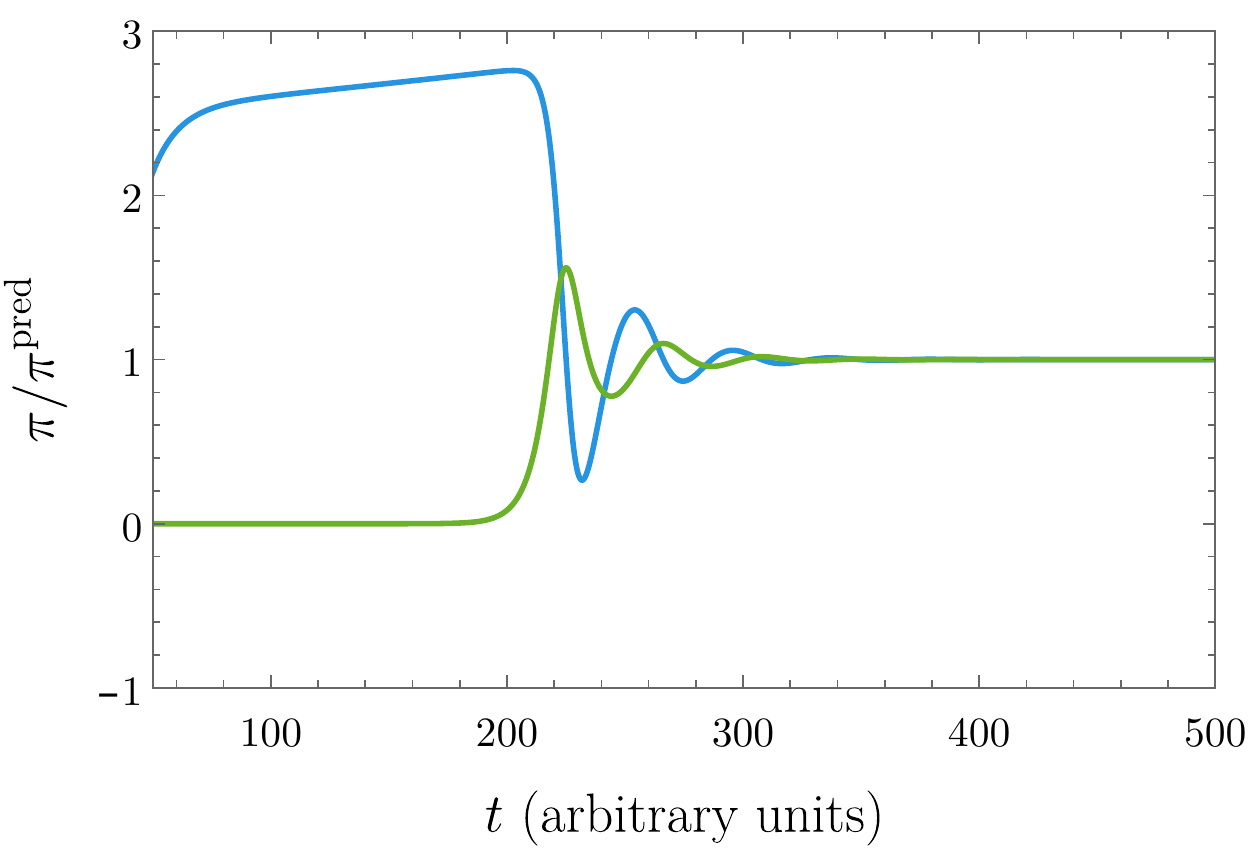}
   % \caption{Fraction of negative eigenvalues}
%    \label{fig:diso}
    \end{subfigure}
%    \caption{These graphs show the effect of varying the }\label{fig:dSHiso}
%\end{figure}
%
%\begin{figure}
%    \centering
~
    \begin{subfigure}{0.48\textwidth}
         \includegraphics[width=1\textwidth]{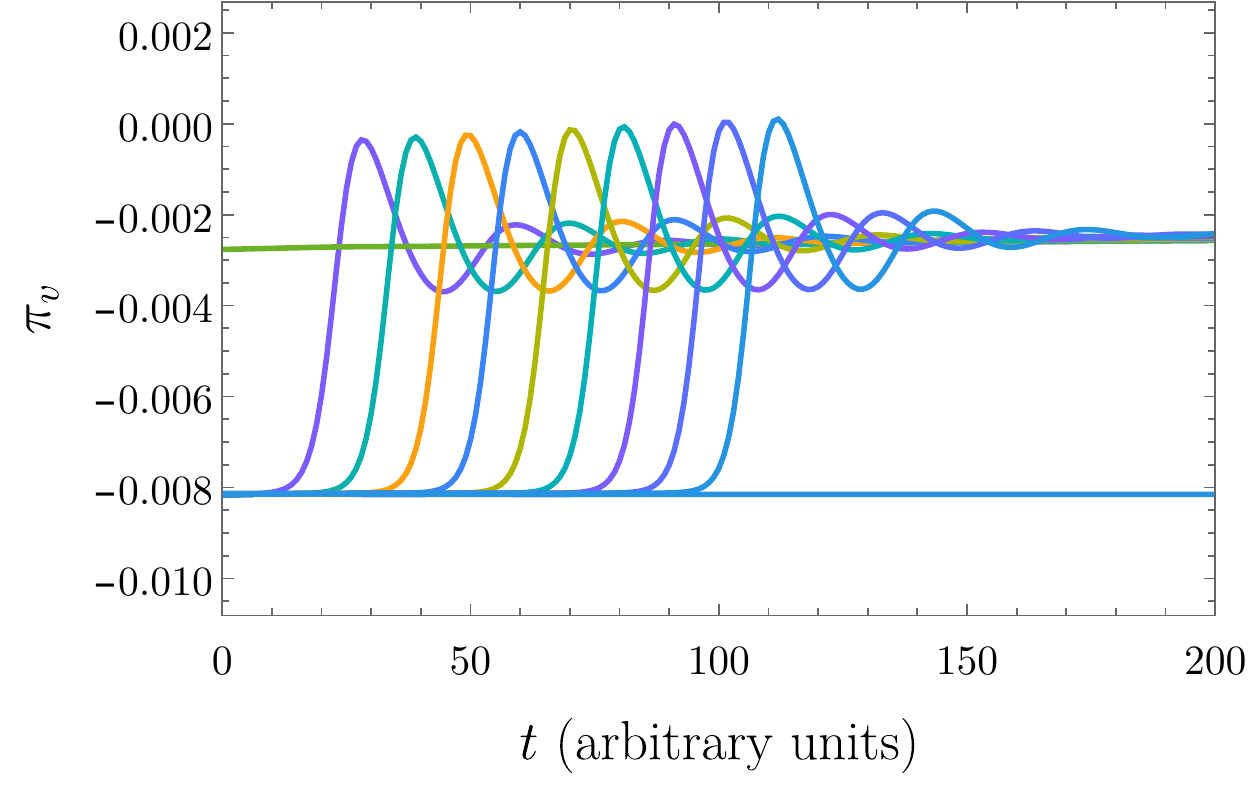}
   % \caption{e-folds before end when a second direction becoms tachyonic}
%    \label{fig:dns}
    \end{subfigure}
     %add desired spacing between images, e. g. ~, \quad, \qquad, \hfill etc. 
      %(or a blank line to force the subfigure onto a new line) \\
	    \caption{Three graphs for the antisymmetric hyperinflation model of Example 3. The top figure shows five trajectories in field space converging to the hyperinflation attractor. The left figure shows how the momenta $\pi_v$ and $\pi_w$ converge to the correct values in one example. The right figure shows how slow-roll with a small velocity in the $w$-direction quickly becomes unstable for $\pi_{w,i}/\pi_{v,i}=10^{-2},\dots,10^{-10}$.\label{fig:nsymmHI2}}
\end{figure}

Figure \ref{fig:nsymmHI2}  (which again have $L=0.05$ and $m=0.01$) displays several aspects of the hyperinflation attractor. The phase space plot (top), 
shows that for all sampled initial conditions (for $\phi_i$, $\theta_i$, $\dot\phi_i$, and $\dot\theta_i$), the momenta $\pi_{v}$ and $\pi_w$ consistently converge to the values predicted by the hyperinflation solution. The left plot shows how $\pi_v$ (blue) and $\pi_w$ (green) converge to the predicted values for one set of initial conditions. The final, right plot shows how trajectories that start in near slow roll (straight blue line) converge to hyperinflation (straight green line). The example shown are inflation models where the initial value for $\pi_v$ is $-V_v/3H$ and $\pi_w = k \pi_v$ where $k$ runs from $10^{-2}$ through $10^{-10}$. As the figure illustrates, the slow-roll solution is exponentially unstable.

\subsection{Example 4: small-field hyperinflation}

So far we have only given examples of large-field models with $\Delta \phi_{\rm tot} >1$. 
Small-field models of  hyperinflation are more intricate to realise, in particular for steep potentials. This example foreshadows some of the issues discussed in the next section. 

We consider the metric  \eqref{eq:Hplane2} and the scalar potential 
\eq{
V=V_\star\left[1+\frac{2\tilde\epsilon}{\tilde\eta}\left(e^{\tilde\eta(\phi-\phi_\star)/3L}-1\right)\right],
}
with positive parameters $\tilde\epsilon$ and $\tilde\eta$. This is a simple way of realising a small-field inflation model with a small $\epsilon$ and negative $\eta$. These are fixed by choosing the values for $\epsilon$ and $\eta$ at $\phi_\star$, giving, $\tilde \epsilon=\epsilon_\star$ and $\tilde\eta=2\epsilon_\star-\eta_\star$, and we will take $\phi_\star$ to correspond to the value of $\phi$ when the CMB modes exited the horizon.

We would now like to construct a model with  $\Delta\phi<1$ and $V_{;v}/V$ as large as possible, that is, a small-field model with a steep gradient. To do this, we set $\epsilon_\star=5.80\times10^{-6}$ and $\eta_\star=-4.28\times10^{-4}$, corresponding to the parameters $\tilde \epsilon=5.80\times10^{-6}$ and $\tilde\eta=4.40\times10^{-4}$, and fix $L=3.87\times 10^{-5}$. The model generates 60 e-folds of inflation over a distance $\Delta\phi=0.2$. It also generates a correctly normalised power spectrum and spectral index ($n_s=0.965$) \cite{Aghanim:2018eyx,Akrami:2018odb}. We note however that the maximal reheating temperature of this model is at the lower edge of what can possibly be compatible with Big Bang Nucleosynthesis: $T_{\rm rh} \lesssim 4\, \text{MeV}$. We will return to this serious issue in section \ref{sec:swampland}.

%(The numerical values  are fixed by minimising $\epsilon$ subject to the constraint that $V_{;v}\geq0.1$, and that reheating can be achieved (more details on this in the next section). 
\subsection{Example 5: multifield hyperinflation}

\begin{figure}
    \centering
    \begin{subfigure}{0.48\textwidth}
         \includegraphics[width=1\textwidth]{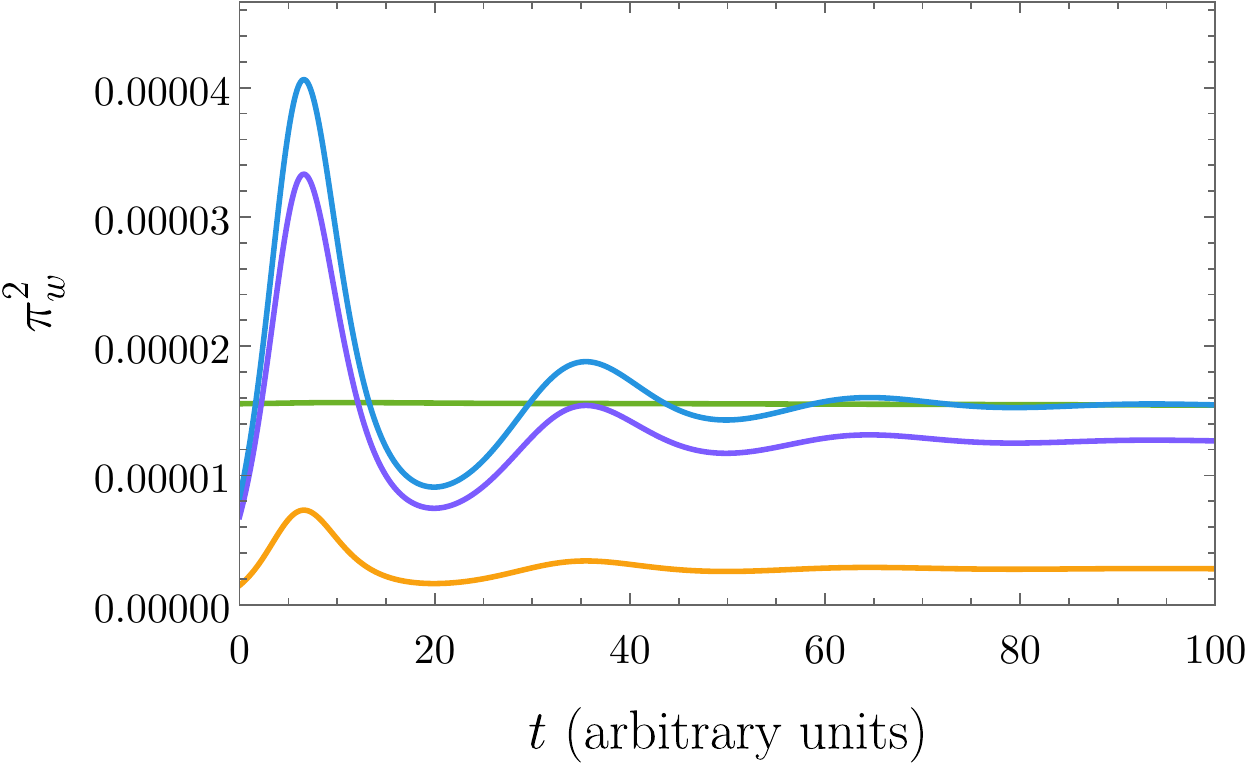}
   % \caption{e-folds before end when a second direction becoms tachyonic}
%    \label{fig:dns}
    \end{subfigure}
     %add desired spacing between images, e. g. ~, \quad, \qquad, \hfill etc. 
      %(or a blank line to force the subfigure onto a new line) \\
	    \caption{The convergence of $\pi^2_{w1}$ (orange), $\pi^2_{w2}$  (purple)  and their sum (blue)    
	    to the analytic hyperinflation solution (green) of Example 5 with $\Nf=3$. \label{fig:3field}}
\end{figure}

Finally, we give an example of a hyperinflation model with more than two fields. This particular model has $\Nf=3$, with a metric and a potential given by,
\eq{
G_{ab}= \text{diag}\left(1,L^2\sinh^2(\phi/L),L^2\sinh^2(\phi/L)\sin^2\theta \right),\hspace{0.5cm}V=\frac12 m^2\phi^2.\label{eq:3field}
}
with the parameter values $L=0.05$, $m=0.01$. This is a straightforward generalisation of the original hyperinflation solution with a quadratic potential, and the numerical solution again agrees very well with the analytic predictions. Figure \ref{fig:3field} shows how the square of the orthogonal momenta converge to the predicted values.

\section{Hyperinflation and the swampland}
\label{sec:swampland}

Various conjectures have recently been proposed to delineate the effective theories that can arise from consistent theories of quantum gravity (the landscape), from those that cannot (the swampland) (see e.g.~\cite{ArkaniHamed:2006dz, Obied:2018sgi, Agrawal:2018own, Danielsson:2018ztv} and references therein). Some of these conjectured conditions have strong implications for early universe cosmology, including models of inflation. For example, a simple form of the `weak gravity conjecture' (WGC) applied to `0-form' gauge potentials \cite{ArkaniHamed:2006dz} limits axion decay constants to $f_a \lesssim 1$ in natural units. If true, this would rule out `natural inflation' in which the inflaton is an axion with a super-Planckian axion decay constant \cite{Freese:1990rb}. 
The `swampland distance conjecture' (cf.~e.g.~\cite{Obied:2018sgi,Ooguri:2018wrx}) states that the maximal field space displacement over which an effective field theory is valid is bounded from above. In some version of this conjecture, the bound is taken to be close to the Planck scale, $\Delta \phi \lesssim 1$ \cite{Agrawal:2018own}.
If true, this conjecture would rule out all EFT descriptions of large-field inflation.
 Moreover, the `swampland de Sitter conjecture' (in its second incarnation  \cite{Ooguri:2018wrx,Garg:2018reu})  states that the norm of the gradient of the potential is bounded from below: $|\nabla V| > {\cal O}(V)$, unless the Hessian matrix has a sufficiently tachyonic eigenvalue. If true, this conjecture would rule out standard single-field slow-roll inflation. The status of these conjectures remain highly uncertain and controversial (cf.~e.g.~\cite{Denef:2018etk,Akrami:2018ylq,Cicoli:2018kdo,Marsh:2018kub}), and in this section we simply explore their consequences for hyperinflation without suggesting their broad validity.

Multifield models with rapidly turning trajectories have been proposed to provide a possible way to circumvent the swampland criteria \cite{Achucarro:2018vey}. 
Hyperinflation provides a concrete and specific class of models of this type, and in this section we critically  discuss hyperinflation in the light of the swampland conjectures. 
 %We show that hyperinflation can satisfy the de Sitter conjecture \emph{or} the distance conjecture, but not both simultaneously while also managing to reheat the universe at the end of inflation. 
 We show that 
 while 
 hyperinflation 
 is less constrained by the swampland conditions than standard slow-roll inflation, it
 cannot simultaneously (1) satisfy the de Sitter conjecture, (2) satisfy the distance conjecture, and (3) manage to reheat the universe to $T\gtrsim 5\, {\rm GeV}$ at the end of inflation.  We furthermore note that the simplest version of hyperinflation is in strong tension with the WGC, but generalised versions of hyperinflation may avoid this conjecture.

\subsection{Hyperinflation and the weak gravity conjecture}
\label{sec:WGC}
The two-field system with a rotationally symmetric potential that we reviewed in section \ref{sec:HI} admits an interpretation as a dilaton/modulus field ($\phi$) coupled to an axion. An immediate consequence of this interpretation is that the axion decay constant is exponentially large during inflation:
\be
f_a \sim L \sinh(\phi/L) \sim \frac{L}{2} e^{180} \, .
\ee
On can show (e.g.~by a slight adaptation of the arguments that we will give in section \ref{sec:fieldexc}) that it is not possible to construct a simple hyperinflation model with only an axion and a modulus that is both observationally consistent and has $f_a \lesssim 1$. 
Consequently, the arguably most promising realisations of hyperinflation are of the generalised kind that we have presented in this paper, which do not feature axions and for which the WGC does not apply.

\subsection{Observational predictions for generalised hyperinflation}
\label{sec:obsgen}
%We close this section by briefly commenting on further restrictions imposed by requiring 
In this section, we briefly discuss the observational consistency of certain sub-classes of realisations of hyperinflation. 
%
%
%To understand the implications of this large turning rate, 
The characteristic properties of the predictions follow directly from the exponential amplification of the  adiabatic perturbations around horizon crossing \cite{Brown:2017osf}.  According to \cite{Mizuno:2017idt}, the power-spectrum of the (constant-density gauge) curvature perturbation, $\zeta= -\dot\phi_a\delta\phi^a/2\epsilon H$ \cite{Dias:2015rca}, is well approximated  by
\eq{
P_\zeta=\frac{H^2}{8\pi^2\epsilon }\gamma(\bar h)^2
\label{eq:Pofzeta}
}
where the function $\gamma(\bar h)^2$ parametrises the enhancement and is (for 
 $\bar h\gtrsim 5$)
given by
\eq{
\gamma(\bar h)^2=\frac{9+\bar h^2}{\bar h^2}e^{2p+2q\bar h} \label{eq:gofh} \, ,
}
with  $p=0.395$ and $q=0.924$.

We begin by showing  that models of hyperinflation with a small inflationary field excursion, $\Delta \phi <1$,
a monotonically increasing $\epsilon$ parameter, and with $\bar h \gtrsim 5$ are ruled out by observations. 
%
%that the spectral index of the primordial perturbations is observationally consistent. 
From equation \eqref{eq:Pofzeta}, it follows that the spectral index is to leading order in the slow-roll parameters given by
\eq{
n_s-1\simeq -2\epsilon +q \bar h\eta \, . %=3(qh-1)\epsilon_L-3qh\eta_L
}

We first consider the case when $\epsilon$ grows monotonically during inflation ($\eta>0$), and $\Delta \phi <1$. In this case the field excursion is bounded  by
\be
1> \Delta \phi > \sqrt{3\epsilon_L^\star} N_{\rm tot} \, ,
\ee
where $\epsilon^{\star}_{L}$ is evaluated at the horizon crossing of the pivot scale. Clearly $\epsilon_L^\star < 1/(3 N_{\rm tot}^2)$ and the negative  $2\epsilon$ contribution to the spectral index is too small to explain the observed deviation from scale invariance. To be compatible with the observed red spectral tilt, the second term ($q\bar h\eta$) needs to be negative. However this requires $\eta<0$, so that $\epsilon$ is not monotonously increasing, contrary to our assumption. Thus, rapidly turning small-field models of hyperinflation  cannot explain the spectral tilt if the first slow-roll parameter is monotonically increasing during inflation.

Finally, we briefly comment on the perturbations in the general hyperinflation case with $\Nf>2$ fields. From equation \eqref{eq:iso}, we see that on superhorizon scales for $\bar h \approx \omega \gg 1$, the (spatially flat-gauge) amplitudes of these fields  grow as $\sqrt{\epsilon}$, just as the adiabatic mode. The superhorizon evolution  therefore does not suppress the entropic modes relative the adiabatic perturbations, and one might  expect these models to produce large amounts of power in the isocurvature modes. This is not the case. The exponential amplification of the adiabatic perturbations reflected in equation \eqref{eq:Pofzeta} has no counterpart for the entropic perturbations, which emerge relatively suppressed on superhorizon scales. The isocurvature to curvature ratio is given by
%, but one also needs to take into account the sub-horizon evolution of the modes. It has been shown that the adiabatic mode sees an exponential increase (in $\omega$) before it crosses the horizon. The entropic mode in the $v^a$-$\tensor{w}{_1^a}$ plane ends up being suppressed on superhorizon scales, and will be irrelevant at the end of inflation. The ratio of power in the isocurvature modes to the curvature mode at the end of inflation will therefore be
\eq{
\frac{P_{\mathcal S}}{P_\zeta}=(\Nf-2)\frac{h^2}{9+h^2}e^{-2p-2qh} \, ,
}
where we made use of results in \cite{Mizuno:2017idt}.

\subsection{Small-field hyperinflation in steep potentials and  reheating}
\label{sec:fieldexc}
Small-field models of hyperinflation satisfy $\Delta \phi <1$ and are compatible with the `swampland distance conjecture'. In this section, we consider the implications of small-field models of hyperinflation in steep potentials satisfying $V_{;v}/V >c$ where $c$ is assumed to be an ${\cal O}(1)$ number. 

Given $\epsilon_L > cL$, the limit on the  total field excursion during inflation implies that, 
\be
1> \Delta \phi_{\rm tot} = \int \sqrt{3 \epsilon_L} dN > \sqrt{3c L}\, N_{\rm tot}  \, .
\ee
We may cast this as an upper limit on the curvature scale: $L < 1/(3c N_{\rm tot}^2)$. The turning rate parameter $\bar h$ is related to $\epsilon_L$ and $L$ by
\be
 \bar h^2 = 3\left(\frac{\epsilon_L}{L^2}-3\right) > 3\left(\frac{c }{L}-3\right) >  9\left(c^2 N_{\rm tot}^2 -1\right) \approx 9 c^2 N_{\rm tot}^2 \gg 1 \, .
 \label{eq:hbarineq}
\ee

From equation \eqref{eq:hbarineq} we see that the exponent of the enhancement factor is bounded from below by a rather large number, and the amplification of the perturbations is enormous. Still, the amplitude of the primordial curvature perturbation at the `pivot scale' of $k_\star=0.05$ Mpc$^{-1}$ is fixed by observations of the cosmic microwave background to $P_\zeta(k_\star) = P_\star= 2.2\times10^{-9}$. As we will now argue, it is challenging to match the amplitude of the primordial power spectrum  in  models of hyperinflation  with $\Delta \phi <1$ and $V_{;v}/V>1$.

%We may now combine equation and \eqref{eq:gofh} for $\bar h\gg 3$ to obtain the limit,

Obtaining the right amplitude of the power spectrum requires,
\be
e^{2q\bar h} =  \epsilon_{\star} e^{-2p} \frac{24\pi^2 P_{\star}}{V_{\star}} <e^{-2p} \frac{24\pi^2 P_{\star}}{V_{\star}} \, ,
\label{eq:Vcond}
\ee
which may be combined with equation \eqref{eq:hbarineq} to give,
\be
N_{\rm tot} < \frac{1}{6qc} \ln \left( \frac{24\pi^2 P_{\star}}{V_{\star}} e^{-2p}\right) = 24.6 \times \frac{1}{c}\left( 1+0.0074\ln\left(\frac{(100\, {\rm GeV})^4}{V_{\star}} \right) \right) \, .
\label{eq:Ntotlimit}
\ee
Thus, these models cannot generate an arbitrarily large number of e-folds of expansion and a consistently normalised power spectrum. Indeed, as we will now show, for $c=1$ and inflationary energy scales $V_\star^{1/4} \gtrsim 4.6\, {\rm GeV}$, these models cannot solve the horizon and flatness problems.

The inequality \eqref{eq:Ntotlimit} rules out most of the possible parameter space for small-field hyperinflation in steep potentials. We here note that a small, remaining window in principle exists, albeit at the expense of fine-tuning.

The inflationary epoch must be followed by a hot big bang cosmology that, in particular, can explain big bang nucleosynthesis and the thermalisation of the neutrinos. This puts a lower limit on the reheating temperature of the universe to $T \geq T_{\rm min} \approx 4$ MeV \cite{deSalas:2015glj}. Additional requirements, such as the generation of the baryon asymmetry, may require a much higher reheating temperature, but are model dependent. Assuming (conservatively) instant reheating at the end of inflation, and that all the energy of the inflaton is transferred into the thermal plasma, we express  the limit on $T$ as a limit on the inflationary potential: $V_\star > V_{\rm end} \geq V_{\rm min} = \tfrac{\pi^2}{30} g_\star(T_{\rm min}) T_{\rm min}^4$, where $g_\star(T_{\rm min}) = 10.75$.  This gives $3H^2 \approx V_\star > V_{\rm min} =(6\, \text{MeV})^4$. We note that achieving rapid reheating requires a strong coupling of the inflaton to the Standard Model, which  makes such models highly sensitive to loop corrections. We will not discuss this issue further here, but simply note that models with inflationary energies close to $V_{\rm min}^{1/4}$ are subject to multiple additional challenges.

%These models are only compatible with observations in the rather extreme limit of a very small energy scale of inflation combined with instantaneous reheating. 
Solving the flatness and horizon problems requires $N_{\rm tot} \gtrsim N_{\rm min} \approx 62 - \ln(10^{16}\, \text{GeV}/V_{\rm end}^{1/4})$ e-folds of expansion if the reheating is instantaneous. Small-field  hyperinflation in steep potentials ($c=1$) provides a solution to these classical problems of big bang cosmology only if
\be
(6\, \text{MeV})^4<V_{\star} \approx V_{\rm end} < (4.6~\text{GeV})^4\, .
\label{eq:Vconstr}
\ee
In other words, the enormous enhancement of the adiabatic perturbations during inflation forces the energy scale of inflation to be very low, pushing it into tension with reheating. 
The inequality \eqref{eq:Vconstr} is a severe, general constraint on these models, and we expect that constructing explicit, realistic  realisations will  be very challenging, though not necessarily impossible.\footnote{For example, baryogenesis may be achieved through the Affleck-Dine mechanism and dark matter may be non-thermally produced, e.g.~as axions.} 

To avoid these constraints altogether in hyperinflation, one must relax either of the conditions $\Delta  \phi <1$ or $V_{;v}/V >c$. 
Our example in section \ref{sec:multifield} demonstrates that observationally consistent small-field models are possible, for example with $V_{;v}/V = 0.1$.

\section{Conclusions}
We have generalised the mechanism of hyperinflation to incorporate broad classes of theories in $\Nf\ge2$ dimensional field spaces with constant negative curvature. In particular, we have shown that hyperinflation does not rely on rotationally symmetric scalar potentials, and can even be realised in theories with randomly interacting fields and without special initial conditions. In some models, hyperinflation can follow a period of slow-roll inflation that becomes  `geometrically destabilised'  by the negative curvature; however, hyperinflation differs from the realisations of `side-tracked inflation' studied in \cite{Garcia-Saenz:2018ifx}.  
 
We have provided the first explicit proof of the attractor nature of hyperinflation and its generalisation, and provided a set of non-trivial explicit examples of the mechanism. These include the first explicit examples without rotational symmetry, with more than two fields, and a small-field model with an observationally consistent power spectrum.  

We have furthermore shown that hyperinflation is in some tension with various `swampland conjectures'. The simplest models are in stark conflict with the weak gravity conjecture, but more general realisations need not be.  Moreover, hyperinflation can be realised in steep potentials but is hard pressed to satisfy both $\Delta \phi <1$ and $|\nabla V|/V >1$: these models give an enormous exponential enhancement of the primordial density perturbations, and can only be realised in models with an extremely low energy-scale of inflation ($6\, \text{MeV} \lesssim V^{1/4} \lesssim 5\, \text{GeV}$). However, we have have also shown that slightly relaxing these conditions, e.g.~by allowing $|\nabla V|/V=0.1$, may permit for the construction of observationally viable models. 
  
%We have confirmed hyperinflation  certain models can be compatible with observational limits at the level of the power spectrum. 

Several extensions of our analysis in this paper are possible. Our results show that (generalised) hyperinflation can be compatible with observations at the level of the power spectrum, however, we have not investigated the possible non-Gaussian signals of the exponentially amplified adiabatic perturbations. Moreover, while our analysis has clarified some challenges for embedding hyperinflation within fundamental physics, explicit realisations within an ultraviolet complete framework are still lacking. Moreover, hyperinflation may be realised in more general field space geometries  in which the curvature scale $L$ varies. Such constructions could possibly avoid our restriction on  steep, small-field hyperinflation. Finally, while we have noted some similarities between hyperinflation and side-tracked inflation \cite{Garcia-Saenz:2018ifx}, the nature of this connection is still poorly understood. Given the prevalence of negatively curved field spaces in high-energy physics models, it would be very interesting to see if there are any common elements to these non-slow-roll inflationary models.\footnote{The nature of this relationship is explained in subsequent work by one of the authors: it can be shown that they are both examples of `rapid-turn attractors' \cite{Bjorkmo:2019fls}.}

%, and whether these could also be found in other systems as well.

\subsection*{Acknowledgements}
We are very grateful to Adam Brown for comments on a draft of this paper, and to Liam McAllister and Sebastien Renaux-Petel for discussions. 
T.B. is funded by an STFC studentship at DAMTP, University of Cambridge. D.M.~gratefully acknowledges past support from a Stephen Hawking Advanced Fellowship at the Centre for Theoretical Cosmology, DAMTP, University of Cambridge.

\appendix

\section{Perturbation theory \label{appendix:stability}}

In this appendix we look at linear perturbations around the hyperinflation attractor, working to first order in the inflationary parameters $\epsilon$ and $\eta$. We do this using the mathematical framework of vielbeins, in the manner of Achucarro et al. \cite{Achucarro:2010da}, and begin by looking at the two-field case. For our purposes it is convenient to write the equations of motion using e-folds as the time-variable, giving
\eq{
\mathcal D_N\mathcal D_N\delta\phi^a+(3-\epsilon)\mathcal D_N\delta\phi^a+[(k^2/a^2H^2)\tensor\delta{^a_b}+\tensor{\tilde M}{^a_b}]\delta\phi^b=0\label{eq:perteq2},
}
where $\tensor{\tilde M}{^a_b}\equiv\tensor{M}{^a_b}/H^2$. 

It is convenient to work with perturbations in a vielbein basis, definied by
\eq{
\delta\phi^I\equiv \tensor{e}{^I_a}\delta\phi^a\hspace{0.5cm}\Leftrightarrow\hspace{0.5cm}\delta\phi^a=\tensor{e}{_I^a}\delta\phi^I.
}
Our basis vectors satisfy $\tensor{e}{^I_a}\tensor{e}{^J_b}G^{ab}=\tensor{\delta}{^I^J}$ and $\tensor{e}{^I_a}\tensor{e}{^J_b}\delta_{IJ}=G_{ab}$. The $I,~J$ indices are raised with deltas. From these identities it also follows that
\eq{
\tensor{e}{^I_a}\mathcal D_N \tensor{e}{_J^a}=-\tensor{e}{_J^a}\mathcal D_N \tensor{e}{^I_a},\hspace{1cm}\tensor{e}{^I_a}\mathcal D_N \tensor{e}{_I^b}=-\tensor{e}{_I^b}\mathcal D_N \tensor{e}{^I_a}.\label{eq:asymmetry}
}
One can introduce a covariant derivative for these perturbations, given by
\eq{
\mathcal D_NX^I\equiv \frac{d X^I}{dN}+\tensor{Y}{^I_J}X^J, 
}
where the antisymmetric matrix $\tensor{Y}{^I_J}$ is defined by
\eq{
\tensor{Y}{^I_J}\equiv\tensor{e}{^I_a}\mathcal D_N\tensor{e}{_J^a}.
}
With this definition this covariant derivative satisfies $\tensor{e}{^I_a}\mathcal D_NX^a=\mathcal D_NX^I$, from which it follows that we can write the equations of motions as
\eq{
\mathcal D_N\mathcal D_N\delta\phi^I+(3-\epsilon)\mathcal D_N\delta\phi^I+[(k^2/a^2H^2)\tensor\delta{^I_J}+\tensor{\tilde M}{^I_J}]\delta\phi^I=0\label{eq:perteq3},
}
where we further defined $\tensor{\tilde M}{^I_J}\equiv\tensor{e}{^J_a} \tensor{\tilde M}{^a_b}\tensor{e}{_J^b}$.

\subsection{Two-field case}

Now we want to write these equations of motion out explicitly for the two-field case. The basis we use for this is the one given by the adiabatic and entropic basis vectors,
\eq{
\tensor{e}{^I_a}=\left(n_a,~s_a\right)^T,\hspace{1cm}\tensor{e}{_I^a}=\left(n^a,~s^a\right).
}
One could also work in the gradient basis defined earlier in the paper, which was very useful in solving the background equation, but for the perturbations this basis makes the results clearer. To change the basis, we need the relations
\eq{
n^a=(\dot\phi_vv^a+\dot\phi_ww^a)/\dot\phi,\hspace{1cm}s^a=(-\dot\phi_wv^a+\dot\phi_vw^a)/\dot\phi,
}
and as an aside we note that these relations can be inverted to
\eq{
v^a=(\dot\phi_vn^a-\dot\phi_ws^a)/\dot\phi,\hspace{1cm}w^a=(\dot\phi_wn^a+\dot\phi_vs^a)/\dot\phi.
}

We begin by looking at the mass matrix in our new basis. Using $R=-2/L^2$ we straightforwardly find
\eq{
\tensor{\tilde M}{^I_J}=\begin{pmatrix}\omega^2-{3\eta}/2&-3\omega(1-\epsilon/3+\eta/2)-9\eta/2\omega\\ -3\omega(1-\epsilon/3+\eta/2)-9\eta/2\omega&-\omega^2(1-2\epsilon/3+\eta/3)-3\eta-27\eta/2\omega^2\end{pmatrix},
}
where we for later convenience has written it in terms of, $\omega$, the rotation rate of the system as defined in \ref{eq:rotation}. It is straightforward to check that this matrix has eigenvalues $\pm \omega\sqrt{9+\omega^2}+\mathcal O(\epsilon)$. This does however not automatically mean that the system is unstable: we also need to take into account the kinetic terms, where we have terms arising from the rotation of the basis vectors.

We now need to compute the matrices  $\tensor{Y}{^I_J}$. By looking at equation \ref{eq:rotation} we see that they are given by
\eq{
\tensor Y{^I_J}=\omega\begin{pmatrix}0&-1\\1&0\end{pmatrix},\label{eq:adbasisrot}
}
and by using $ d\omega/dN=(9+\omega^2)\eta /2\omega+\mathcal O(\epsilon^2)$, we find the complete equations of motion
\begin{align}
\frac{d^2\delta\phi^n}{dN^2}+(3-\epsilon)\frac{d\delta\phi^n}{dN}-2\omega\frac{d\delta\phi^s}{dN}+\left[\frac{k^2}{a^2H^2}-\frac{3\eta}2\right]\delta\phi^n+\left[-\omega(6-2\epsilon+\eta)-\frac{(9+\omega^2)\eta}\omega\right]\delta\phi^s&=0\\
\frac{d^2\delta\phi^s}{dN^2}+(3-\epsilon)\frac{d\delta\phi^s}{dN}+2\omega\frac{d\delta\phi^n}{dN}-\omega\eta\delta\phi^n+\left[\frac{k^2}{a^2H^2}-\frac13\omega^2(6-2\epsilon+\eta)-\frac{3(9+\omega^2)\eta}{\omega^2}\right]\delta\phi^s&=0\,.
\end{align}
To see that this implies that the background equations are linearly stable we just need to find the four solution for $k=0$ (taking $\epsilon$ and $\eta$ to be constant). Defining $\delta\pi^I\equiv d\delta\phi^I/dN$, we can rewrite the system as
\eq{
\frac{d}{dN}\begin{pmatrix}\delta\phi_n \\ \delta\phi_s \\ \delta\pi_n \\ \delta\pi_s\end{pmatrix}=
\begin{pmatrix}0 & 0 & 1 &0\\
0 & 0 & 0 &1\\
3\eta/2, & \omega(6-2\epsilon+\eta)+\frac{(9+\omega^2)\eta}\omega ,&-(3-\epsilon) & 2\omega\\
\omega\eta, & \frac13\omega^2(6-2\epsilon+\eta)+\frac{3(9+\omega^2)\eta}{\omega^2},& -2\omega & -(3-\epsilon)
\end{pmatrix}
 \begin{pmatrix}\delta\phi_n \\ \delta\phi_s \\ \delta\pi_n \\ \delta\pi_s\end{pmatrix}\equiv U\begin{pmatrix}\delta\phi_n \\ \delta\phi_s \\ \delta\pi_n \\ \delta\pi_s\end{pmatrix}.
}
In this formulation, we can infer whether the system is locally stable by computing the local Lyapunov exponents, i.e. the eigenvalues $\lambda$ of $U$. To first order, they are given by
\eq{
\lambda=\eta/2+\mathcal O(\epsilon^2),~-3+\mathcal O(\epsilon),~\frac12\left(-3\pm\sqrt{9-8\omega^2}\right)+\mathcal O(\epsilon).
}
The latter three all have negative real parts, and thus correspond to decaying modes. The first one is a slowly growing mode corresponding to the direction $(1,0,\eta/2,0)^T$. This of course is just our regular superhorizon scale adiabatic perturbation, $\delta\phi^I\propto(\sqrt\epsilon,~0)$, and corresponds to a shift along the trajectory. All other solutions decaying, we see that the system indeed is stable.

\subsection{Multifield extension}

The extension to more than two fields is reasonably straightforward. From equation \ref{eq:mfvw1}, we see that the matrix $\tensor Y{^I_J}$ is block diagonal,
\eq{
\tensor Y{^I_J}=\begin{pmatrix}Y_{(2)} & 0\\0 & Y_{(\Nf-2)}\end{pmatrix},
}
where $Y_{(2)}$ is our previous $Y$ matrix. We cannot know the form of $Y_{(\Nf-2)}$ without specifying the other basis vectors, however as we shall see, we will not need to make an explicit choice for this. Recalling that in an orthonormal basis $\tensor{R}{^I_{IJKL}}=-(\tensor{\delta}{^I_K}\delta_{JL}-\tensor{\delta}{^I_L}\delta_{JK})/L^2$, we see that the mass matrix is also block diagonal,
\eq{
\tensor M{^I_J}=\begin{pmatrix}M_{(2)} & 0\\0 & M_{(\Nf-2)}\end{pmatrix},
}
where $M_{(2)}$ is our old mass matrix. $M_{(\Nf-2)}$ is a diagonal matrix given by
\eq{
M_{(\Nf-2)}=\left(\frac{V_{;v}}{L}-\frac{\dot\phi^2}{L^2}\right)I_{(\Nf-2)}.
}
Since both the $Y$ and $M$ matrices are block-diagonal, the additional $\Nf-2$ perturbations decouple from the first two at quadratic order. We can therefore solve for their evolution independently of the first two.

We now define a new set of fields $\delta\phi'^I$ (for $I> 2$) through $\delta\phi^I=\tensor{R}{^I_J}\delta\phi'^J$, where the rotation matrix $\tensor R{^I_J}$ satisfies $d\tensor R{^I_J}/dt=-H\tensor Y{_{(\Nf-2)}^I_K}\tensor R{^K_J}$. These fields satisfy \cite{Achucarro:2010da}
\eq{
\mathcal D_t \delta\phi^I=\tensor R{^I_J}\frac{d\delta\phi'^J}{dt},\hspace{1cm}\mathcal D_t\mathcal D_t \delta\phi^I=\tensor R{^I_J}\frac{d^2\delta\phi'^J}{dt^2},
}
giving the equations of motion
\eq{
\frac{d^2\delta\phi'^I}{dN^2}+(3-\epsilon)\frac{d\delta\phi'^I}{dt}+\left[\frac{k^2}{a^2H^2}\tensor{\delta}{^I_J}+\tensor{(R^TM_{(\Nf-2)}R)}{^I_J}\right]\delta\phi'^J=0,
}
but since the mass matrix is diagonal this just reduces to
\eq{
\frac{d^2\delta\phi'^I}{dN^2}+(3-\epsilon)\frac{d\delta\phi'^I}{dt}+\left[\frac{k^2}{a^2H^2}\tensor{\delta}{^I_J}+\tensor{M}{_{(\Nf-2)}^I_J}\right]\delta\phi'^J=0.
}
These are just $\Nf-2$ uncoupled modes, each satisfying
\eq{
\frac{d^2\delta\phi^{s_I}}{dN^2}+(3-\epsilon)\frac{d\delta\phi^{s_I}}{dN}+\left[\frac{k^2}{a^2H^2}-\frac{3(9+\omega^2)\eta}{2\omega^2}\right]\delta\phi^{s_I}=0.
}

\bibliographystyle{JHEP}%plain
\bibliography{HIrefs}

%\begin{thebibliography}{99}

%\bibitem{a}
%Author, \emph{Title}, \emph{J. Abbrev.} {\bf vol} (year) pg.
%
%\bibitem{b}
%Author, \emph{Title},
%arxiv:1234.5678.
%
%\bibitem{c}
%Author, \emph{Title},
%Publisher (year).

% Please avoid comments such as "For a review'', "For some examples",
% "and references therein" or move them in the text. In general,
% please leave only references in the bibliography and move all
% accessory text in footnotes.

% Also, please have only one work for each \bibitem.

%\end{thebibliography}
\end{document}